\documentclass[manuscript,nonacm]{acmart}

\usepackage{braket, dsfont, placeins}
\usepackage[normalem]{ulem}
\usepackage{xspace}
\usepackage{parskip, indentfirst}

\newcommand{\xrm}{\text{x}}
\newcommand{\BV}{Bernstein-Vazirani}
\graphicspath{{figures/}}

\AtBeginDocument{%
  \providecommand\BibTeX{{%
    \normalfont B\kern-0.5em{\scshape i\kern-0.25em b}\kern-0.8em\TeX}}}

\setcopyright{acmcopyright}
\copyrightyear{2023}
\acmYear{2024}
\acmDOI{TBD}

\begin{document}

\title{Impact of unreliable devices on stability of quantum computations}

\author{Samudra Dasgupta$^{1,2}$}
\email{sdasgup3@tennessee.edu}
\orcid{0000-0002-7831-745X}
\author{Travis S. Humble$^{1,2}$}
\orcid{0000-0002-9449-0498}
\email{humblets@ornl.gov}
\affiliation{%
  \institution{$^1$Quantum Science Center, Oak Ridge National Laboratory, Oak Ridge, Tennessee; $^2$Bredesen Center, University of Tennessee, Knoxville, Tennessee}
  \country{USA}
}

\renewcommand{\shortauthors}{Dasgupta and Humble}

\begin{abstract}
Noisy intermediate-scale quantum (NISQ) devices are valuable platforms for testing the tenets of quantum computing, but these devices are susceptible to errors arising from de-coherence, leakage, cross-talk and other sources of noise. 
This raises concerns regarding the stability of results when using NISQ devices since strategies for mitigating errors generally require well-characterized and stationary error models. 
Here, we quantify the reliability of NISQ devices by assessing the necessary conditions for generating stable results within a given tolerance. 
We use similarity metrics derived from device characterization data to derive and validate bounds on the stability of a 5-qubit implementation of the \BV~algorithm. 
Simulation experiments conducted with noise data from IBM washington, spanning January 2022 to April 2023, revealed that the reliability metric fluctuated between 41\% and 92\%. This variation significantly surpasses the maximum allowable threshold of 2.2\% needed for stable outcomes. Consequently, the device proved unreliable for consistently reproducing the statistical mean in the context of the \BV~circuit.
\end{abstract}



\received{30 May, 2023}
\received[revised]{{[15 May, 2024]}}
\received[accepted]{{[30 June, 2024]}}

\maketitle

\textit{
This manuscript has been authored by UT-Battelle, LLC under Contract No. DE-AC05-00OR22725 with the U.S. Department of Energy. The United States Government retains and the publisher, by accepting the article for publication, acknowledges that the United States Government retains a non-exclusive, paid-up, irrevocable, worldwide license to publish or reproduce the published form of this manuscript, or allow others to do so, for United States Government purposes. The Department of Energy will provide public access to these results of federally sponsored research in accordance with the DOE Public Access Plan (https://www.energy.gov/doe-public-access-plan).
}


\clearpage
\section{Introduction}\label{sec:introduction}
\begin{figure}
  \centering
\includegraphics[width=0.5\textwidth]{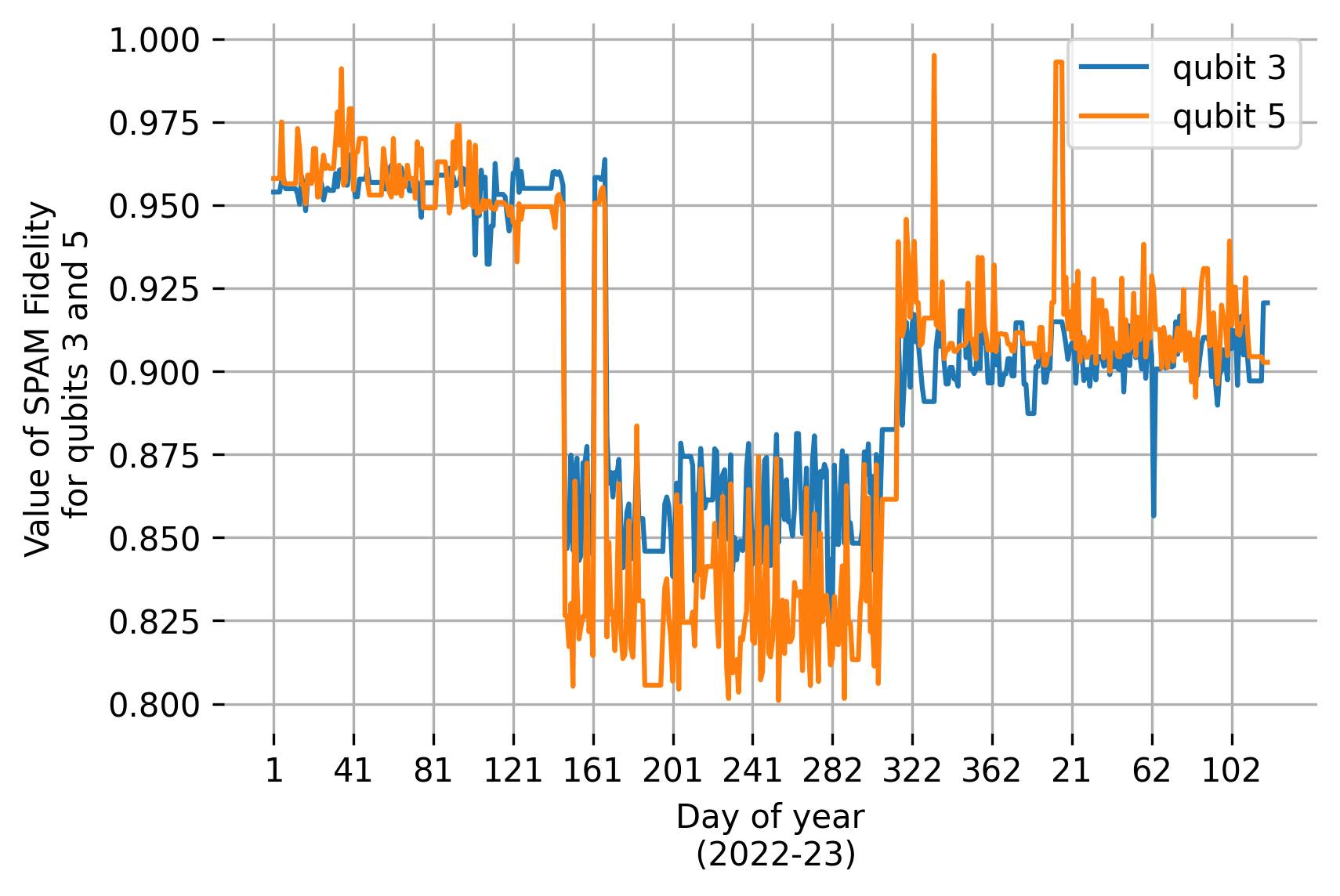}
\caption{Daily estimates of the state preparation and measurement (SPAM) fidelity for two register elements in a superconducting device collected {from 1-Jan-2022 to 30-Apr-2023}. Day of the year is indexed starting with 1 and index 365 corresponds to Dec 31.}
\Description{Daily estimates of the state preparation and measurement fidelity for two register elements in a superconducting  device.}
\label{fig:q_3_5_washington_jan2022_apr2023}
\end{figure}
Ongoing efforts to realize the principles of quantum computing have demonstrated control over quantum physical systems ranging from superconducting electronics \cite{burnett2019decoherence}, trapped ions \cite{wan2019quantum}, and silicon quantum dots \cite{PhysRevApplied.10.044017} among many others \cite{humble2019quantum}. As these efforts aim for future fault-tolerant operation \cite{gottesman1998theory}, they have established a regime of noisy, intermediate-scale quantum (NISQ) devices that provide a frontier for testing the principles of quantum computing under experimental conditions \cite{preskill2019quantum}. Such first-in-kind NISQ computing devices enable design verification \cite{PhysRevA.101.042315}, device characterization \cite{harper2020efficient}, program validation \cite{ferracin2019accrediting}, and a breadth of testing and evaluation for application performance \cite{kandala2017hardware,dumitrescu2018cloud,hempel2018quantum,klco2018quantum,mccaskey2019quantum,roggero2020quantum} with several recent demonstrations exemplifying the milestone of quantum computational advantage \cite{arute2019,google2020hartree,zhong2020quantum}.

Practical efforts to realize a quantum computer encounter multiple sources of noise \cite{kliesch2021theory, ferracin2021experimental, coveney2021reliability, blume2010optimal}. 
For example, the physical realization of the quantum register is subject to noise processes (such as decay and de-coherence, coupling to the environment, intra-register cross-talk, and computational leakage) which reduce the lifetime for storing quantum information. 
In addition, the physical implementation of quantum logic operations (such as gates and measurements) rely on analog fields subject to pulse distortion, attenuation, and drift, which reduce gate fidelity. 
Furthermore, fluctuations in  thermodynamic controls (such as cryogenic cooling and magnetic shielding) can also introduce noise.

A prominent feature of experimental NISQ computing is that noise limits performance \cite{martinis2015qubit} such that a device must be frequently calibrated to maintain high-fidelity operations.  Many different parameters are used for monitoring noise within NISQ devices and informing benchmarking methods that assess the accuracy of noisy quantum computation \cite{eisert2020quantum, dahlhauser2021modeling}. These parameters are also essential for mitigating noise-induced errors  \cite{temme2017error,kandala2019error,rudinger2019probing,geller2020rigorous,wilson2020just,hamilton2020error}, \cite{PhysRevLett.78.3971, PhysRevA.94.032321}. Error-mitigated benchmarks establish bounds on the statistical significance of experimental demonstrations using NISQ devices and help clarify the conditions under which results are reproducible \cite{proctor2022measuring}. 

There are different temporal and spatial scales on which to describe the noise sources affecting the fidelity of quantum computing devices. Noise sources that act on the time scale of a single gate operation notably create errors that are observed during an individual quantum circuit execution, while variations in the noise process over minutes and days influence experimental reproducibility \cite{mcrae2021reproducible, burnett2019decoherence, schultz2022impact, etxezarreta2021time, demarti2022performance, gulacsi2023smoking}. For example, Fig.~\ref{fig:q_3_5_washington_jan2022_apr2023} shows an example of  daily fluctuations in the estimated state preparation and measurement (SPAM) fidelity from two register elements in a superconducting device for a 16-month period \cite{washington}. Changes in key noise parameters arise from various factors including unintentional drift in the control system as well as intentional changes in how the device is operated \cite{proctor2020detecting, van_Enk_2013, PhysRevApplied.15.014023, PhysRevLett.78.3971}, and how to manage and mitigate fluctuations across these longer scales is an outstanding concern.

In experiments utilizing NISQ devices, it is often necessary to estimate circuit outcomes using a large number of samples, ranging from thousands to millions, and these experiments can span minutes to hours to days depending on the technology and application. While error mitigation methods exist to handle the stationary noise processes assumed during these executions, there is a gap in methods available to account for non-stationary noise processes. Hardware-aware compilation methods \cite{kandala2017hardware} can be used to mitigate some slower fluctuations, but this fails to address whether an observed result can be reproduced elsewhere or by others. Stability is a serious concern for current quantum benchmarking methods, which do not account for the transient nature of device noise nor provide a means to assess how these fluctuations impact the results \cite{etxezarreta2021time}. In order to ensure the stability of benchmarking results across changes in the underlying devices, the reliability of the device itself must be assessed.

{Assessing the reliability of NISQ devices complements concurrent efforts to evaluate the accuracy and stability of noisy quantum computations.} Accuracy is used widely to quantify how well an instance of an experimentally observed output from noisy circuit execution matches a known expected outcome \cite{ferracin2021experimental, kliesch2021theory, bravyi2021mitigating, blume2020modeling, dahlhauser2021modeling, maciejewski2021modeling, geller2020rigorous, hamilton2020error, hamilton2020scalable}, while stability measures whether the program behavior stays bounded in the presence of noise fluctuations \cite{dasgupta2022assessing}. By comparison, reliability quantifies the temporal consistency of characteristic noise. 

{Quantification of device reliability for quantum computing was first introduced in \cite{dasgupta2021stability}. This study builds upon that foundation but differs significantly in the following two respects. Firstly, this study explores the impact of unreliable devices on quantum program outcomes, developing a theoretical framework to estimate an upper bound on outcome stability. The theory links device reliability quantified using Hellinger distance to outcome stability and is validated with synthetic and experimental noise data from a superconducting device. Secondly, this study quantifies the reliability of a composite circuit with multiple noise sources, which has not been previously explored. It addresses the computational challenges of estimating the Hellinger distance using samples from a high-dimensional noise distribution and proposes clustering-based workarounds for Monte Carlo convergence. 
}

In Sec.~\ref{sec:theory}, we formalize device reliability as a measure of consistent performance with time, and establish bounds on stability of outcomes from quantum programs executed on unreliable devices. In Sec.~\ref{sec:bound_verification}, we verify the predicted bounds on stability using a 5-qubit implementation of the \BV~algorithm~\cite{bernstein1993quantum}. Sec.~\ref{sec:conc} concludes with a discussion on key takeaways and limitations.
\section{Theory}\label{sec:theory}
\subsection{Quantifying device reliability}
In this section, we quantify the reliability of noisy quantum devices 
using the distance ($H$) between the distribution of noise (represented by a $d$-dimensional noise parameter vector $\xrm = (\xrm_1, \cdots, \xrm_d)$) at diferent times. 
Specifically, we use the Hellinger distance given by:
\begin{equation}
H = \sqrt{1-\int\limits_{\xrm} \sqrt{f(\xrm; t_1) f(\xrm; t_2)} d\xrm},
\label{eq:hellinger_unmodified}
\end{equation} 
where $f(\xrm;t)$ denote the probability density function for $\xrm$ at time $t$. The Hellinger distance is bounded between 0 and 1 and symmetric with respect to the inputs. It vanishes for identical distributions and approaches unity for distributions with no overlap. We call a device $\varepsilon$-reliable when $H < \varepsilon$. 

Distance measures (like Hellinger distance) suffer from the curse of dimensionality which has the effect that small changes in a distribution yield large changes in the distance value when the numer of dimensions is high (see Appendix~\ref{sec:dimcurse}). An alternate that does not suffer from the curse of dimensionality is $H_{\text{avg}}$, defined as the average over the distances ($H_k$) between the univariate marginal distributions:
\begin{equation}
H_{\text{avg}} = \frac{1}{d} \sum\limits_{k=1}^d H_k.
\label{eq:hellinger_avg}
\end{equation}
A third option is to normalize the measure by exponentiating the integral with the inverse of the dimension:
\begin{equation}
H_\text{normalized} = \sqrt{1-
\left( \int\limits_{\xrm} \sqrt{f(\xrm; t_1) f(\xrm; t_2)}d\xrm \right)^{1/d}
}.
\label{eq:hellinger_normalized}
\end{equation}
Although neither Eqn.~\ref{eq:hellinger_avg} nor Eqn.~\ref{eq:hellinger_normalized} is affected by the curse of dimensionality, Eqn.~\ref{eq:hellinger_normalized} is preferred because it preserves the correlation information between noise parameters, unlike Eqn.~\ref{eq:hellinger_avg}.
\subsection{Bounding outcome stability}\label{sec:impact}

We defined reliability as a measure of the similarity between time-varying distributions of noise parameters. 
In this section, we consider how to obtain bounds on the stability of program outcomes using this similarity measure. 
Outcome stability~\cite{dasgupta2021stability} is defined as the absolute difference in the mean of the observable obtained from the noisy device at times $t_1$ and $t_2$:
\begin{equation}
s(t_1, t_2) = | \braket{\mathcal{O}}_\xrm(t_1) - \braket{\mathcal{O}}_\xrm(t_2) |
\label{eq:s_def}
\end{equation}
where $\braket{\mathcal{O}}_\xrm(t)$ is the average of the noisy quantum observable $\braket{\mathcal{O}_{ \xrm}}$ with respect to the time-varying noise distribution $f_X(\xrm; t)$:
\begin{equation}
\braket{\mathcal{O}}_\xrm(t) = \int\limits \braket{\mathcal{O}_{ \xrm}}  f_X(\xrm; t) d\xrm.
\end{equation}
The subscript $\xrm$ in the quantum observable $\mathcal{O}$ denotes a specific realization of the device noise when sampled from its distribution $f_X(\xrm; t)$. 

As shown in Appendix~\ref{sec:upperbound}, an upper bound on the stability metric can be obtained as:
\begin{equation}
\begin{split}
s_{\text{max}} =& 2c\sqrt{1-(1-H_\text{normalized}^2)^{2d}}, \;\;\;\; c=\underset{\xrm}{\text{sup}} |\braket{O_{ \xrm}}|
\label{eq:smax_normalized}
\end{split}
\end{equation}
Eqn.~\ref{eq:smax_normalized} upper bounds the temporal variations in the expected mean of the circuit outcome using the Hellinger distance. 
\section{Bound verification}\label{sec:bound_verification}
\subsection{Application example}
\begin{figure}[!h]
\centering
\includegraphics[width=0.6\textwidth]{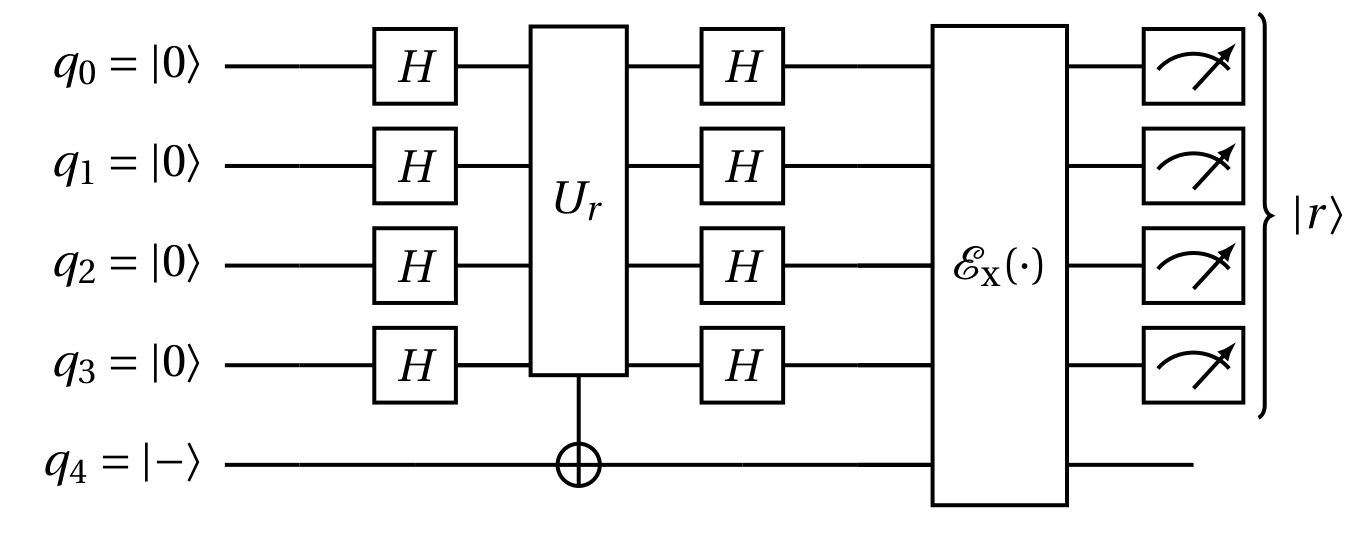}
\caption{
A quantum circuit implementation of the \BV~algorithm that employs 5 qubits, denoted $q_0$ to $q_4$. The first four qubits are used to compute the 4-bit secret string, while the fifth qubit serves as an ancilla and initially resides in the $\ket{-}$ superposition state. The symbol $H$ denotes the Hadamard gate while the oracle unitary ($U_r$) implements the secret string ($r$). The depolarizing noise channel is denoted by $\mathcal{E}_\xrm(\cdot)$. A quantum measurement operation is represented by the meter box symbol at the circuit's end.
}
\Description{
Quantum circuit for the implementation of the \BV~algorithm.
}
\label{fig:bv_ckt_a}
\end{figure}

Our example evaluates the bound on a noisy circuit implementation of the \BV~algorithm \cite{bernstein1993quantum}. This algorithm returns the secret $n$-bit string $r$ encoded into an oracle 
\begin{equation}
f_{\text{BV}}( z ) = z \cdot r,
\end{equation}
by making as few queries as possible. 
The best classical method requires $n$ queries of the oracle while exactly one query is required for the \BV~algorithm. The quantum circuit for a 4-bit secret string is shown in Fig.~\ref{fig:bv_ckt_a} in which $U_r$ implements the oracle unitary. We are interested in the probability of success Pr(r) to compute the secret bit-string $r$ where $\ket{r} = \bigotimes\limits_{i=1}^n \ket{r_i}$ with $r_i \in \{0,1\}$. This probability is given by:
\begin{equation}
\text{Pr(r) = Tr}\left[ \mathcal{O} \rho_{\text{out}}^{\text{noisy}} \right]
\label{eq:prob_r}
\end{equation}
where $\rho_{\text{out}}^{\text{noisy}}$ is the noisy output density matrix and $\mathcal{O} = \ket{r}\bra{r}$  is the observable for the problem.
%
\subsection{Verification using synthetic noise}
In this section, we verify the bound on outcome stability, as shown in Eqn.~\ref{eq:smax_normalized}, using numerical simulations with time-varying depolarizing noise. 

For simplicity, our noise model assumes that each register element is acted upon by isotropic depolarizing noise. We denote this by:
\begin{equation}
\begin{split}
\mathcal{E}_{\xrm_i}(\rho) &= \left( 1-\frac{3\xrm_i}{4} \right)\rho + \frac{\xrm_i}{4} (\mathds{X}_i\rho \mathds{X}_i + \mathds{Y}_i\rho \mathds{Y}_i + \mathds{Z}_i\rho \mathds{Z}_i) \;\;\;\; i \in \{0, \cdots, n-1\}
\end{split}
\end{equation}
where $\mathds{X}_i, \mathds{Y}_i, \mathds{Z}_i$ are the Pauli operators acting on the $i$-th qubit, $\xrm_i$ is the depolarizing parameter for the i-th qubit's noise channel 
and $\rho$ is a n-qubit density matrix. 


The observable $\braket{O_{ \xrm}}$ for a specific realization of circuit noise $\xrm$ is given by:
\begin{equation}
\braket{O_{ \xrm}} =\prod_{i=1}^{n} \left( 1-\frac{\xrm_i}{2}\right).
\label{eq:bv_ox_def}
\end{equation}
The average of the observable over the noise distribution observed at time $t$ is given by: 
\begin{equation}
\braket{O_{ \xrm}}(t) = \int\limits_{\xrm_1, \cdots, \xrm_n} \prod_{i=1}^{n} \left( 1-\frac{\xrm_i}{2}\right) f(\xrm_1, \cdots, \xrm_n; t)
d\xrm_1 \cdots \xrm_n
\end{equation}
Next, we will explore the modeling of the noise distribution $f(\xrm_1, \cdots, \xrm_n; t)$.

\subsubsection{Synthetic dataset} 
Consider a non-stationary noise evolution in between device calibrations 
in which the mean of the de-polarizing noise stays constant with time
(i.e. $\mathds{E}(\xrm_i)   = \mu_0$ $\forall i, t$) 
but the variance steadily increases. 
In particular, let $\omega = \sigma^2_T / \sigma^2_0$ denote the factor by which the variance increases between the two calibration instants $0$ and $T$. Assume that $\mu_0, \sigma_0$ and $\omega$ are known. 
Naively, one might think that the quantum observable does not get impacted in the mean but its variance will increase. 
But that is not the case. 
We will show how the observable starts drifting with time leading to unstable outcomes. 

To model the probability density of the depolarizing parameter $\xrm_i$ for the i-th qubit, we use a beta distribution:
\begin{equation}
f_{X_i}(\xrm_i; t) = \frac{\xrm_i^{\alpha_t-1}(1-\xrm_i)^{\beta_t-1}}{\text{Beta}(\alpha_t, \beta_t)}, 0 \leq \xrm_i \leq 1,
\label{eq:beta_dist}
\end{equation}
where the Beta function, by definition, is:
\begin{equation}
\text{Beta}(m, p) = \int\limits_0^1 y^{m-1}(1-y)^{p-1}dy, \;\;\;\; y\in[0,1].
\end{equation}
The beta distribution is appropriate when the observed data lies between 0 and 1, and exhibits a unimodal skewed histogram, as observed from experimental data seen in Fig.~\ref{fig:f0f1_toronto_qubit_0_onwards_spruce_2021}.

If we assume the functional form for $\alpha_t$ and $\beta_t$ as $\alpha_t = \alpha_0/ (k_0+t)$ and $\beta_t = \beta_0/ (k_0+t)$, then the mean $(\alpha_t/(\alpha_t+\beta_t))$ is independent of time while the variance ($\alpha_t\beta_t/(\alpha_t+\beta_t)^2/(1+\alpha_t+\beta_t)$) increases with time, as desired. The model hyper-parameters $\alpha_0, \beta_0$ and $k_0$  can be derived in terms of $\mu_0, \sigma_0$ and $\omega$ as:
\begin{equation}
\begin{split}
\alpha_0 =& \mu_0 \left[ \frac{\mu_0(1-\mu_0)}{\sigma_0^2} -1\right]\\
\beta_0 =& (1-\mu_0) \left[ \frac{\mu_0(1-\mu_0)}{\sigma_0^2} -1\right]\\
k_0 =& \frac{\phi-1}{\phi/\omega-1}-T \text{ where } \phi = \frac{\mu_0(1-\mu_0)}{\sigma_0^2}\\
\end{split}
\end{equation}
as shown in Appendix~\ref{sec:beta_hyper_params}. Our model further assumes that the depolarizing noise parameters exhibit correlations between them. The i-th row, j-th column element of the correlation matrix $\Sigma$ (i.e. $\Sigma_{ij}$) represents the correlation between the depolarizing parameter acting on qubit $i$ and that acting on qubit $j$.

Having discussed the modeling of the univariate marginals, we now turn to the construction of the joint distribution, for which
we use a copula structure~\cite{sklar1959fonctions}:
\begin{equation}
f_{X}(\xrm;t) = \Theta \left[ F_{X_1}(\xrm_1; t), \cdots F_{X_{n}}(\xrm_{n}; t) \right]
\prod\limits_{j=1}^{n} f_{X_j}(\xrm_j; t).
\label{eq:copulas}
\end{equation}
Here $f_X(\xrm;t)$ is the joint density, 
$F_{X_i}(\xrm_i; t)$ is the cumulative distribution function for the univariate random variable $X_i$, 
and $\Theta(\cdot)$ is the Gaussian copula function:
\begin{equation}
\Theta(y) = \Theta(y_{1},\cdots, y_{n}) = \frac{\exp\left( -\frac{1}{2}(y-\mu_y)^T \Sigma^{-1} (y-\mu_y)\right)}{(2\pi)^{n/2}|\Sigma|^{1/2}},
\end{equation}
which is the standard multi-variate normal distribution with correlation matrix $\Sigma$. 
We chose the Gaussian copula for its simplicity. There exist other copula models such as elliptical copulas and Archimedean copulas which are more suitable for other applications~\cite{zhu2022generative, de2012multivariate, nelsen2007introduction, mcneil2015quantitative, wilkens2023quantum, genest2011inference}. 

Fig.~\ref{fig:depol_sbysmax_large_register} shows how the outcome stability for the \BV~problem deteriorates with the length of the secret-string for various degrees of inter-qubit correlation and has the form: 
\begin{equation}
\begin{split}
s(t_1, t_2; n) \sim&  \sqrt{2\pi n}(e/n)^n\left| 
\sum\limits_{m=0}^M
\frac{1}{m! (n-2m)!}
\left(
{ \theta_{t_1}^m}
\left[1-\frac{\mu_{t_1}}{2}\right]^n
- 
{ \theta_{t_2}^m}
\left[1-\frac{\mu_{t_2}}{2}\right]^n 
\right)
\right|  
\\
\end{split}
\end{equation}
where $\theta_t = [\sigma_t \rho_t / (1-\mu_t/2)]^2/8$ and $\sigma_t, \rho_t, \mu_t$ are the time-varying standard-deviation, correlation coefficient and mean of the de-polarizing parameter $\xrm$. The derivation is provided in Appendix~\ref{sec:scalability}. When this expression becomes uncomputable for large $n$, the upper bound given by Eqn.~\ref{eq:smax_normalized} can be used.
\subsubsection{Distance estimation}\label{sec:distance_estimation}
Next, we estimate the Hellinger distance between the noise distributions at time $t$ relative to the experimentally observed distribution at time $0$. 
Using Eqn.~\ref{eq:beta_dist} to characterize the univariate marginals, the distance can be shown to be:
\begin{equation}
H(t_1, t_2) = \sqrt{1 - \frac{\text{Beta}\left( 
\frac{\alpha_0}{2}\left( \frac{1}{k_0+t_1}+\frac{1}{k_0+t_2} \right), 
\frac{\beta_0}{2} \left( \frac{1}{k_0+t_1}+\frac{1}{k_0+t_2} \right)
\right)}{\sqrt{\text{Beta}(\frac{\alpha_0}{k_0+t_1}, \frac{\beta_0}{k_0+t_1} )}\sqrt{\text{Beta}( \frac{\alpha_0}{k_0+t_2}, \frac{\beta_0}{k_0+t_2})}}}.
\end{equation}
The distance between time-varying multi-variate correlated distribution modeled using Eqn.~\ref{eq:copulas} is analytically intractable. However, it can be estimated using Monte Carlo methods. 
If we draw $N$ samples from the $n$-dimensional distribution $f_{X}(\xrm; t_1)$ (note that $d=n$ in this section) and label the sample set as $\{\xrm^j\}_{j=1}^N$, then we can numerically integrate the integral in Eqn.~\ref{eq:hellinger_unmodified} using:
\begin{equation}
\frac{1}{N}\sum\limits_{j=1}^{N} \sqrt{ \frac{f_{X} (\xrm^j; t_2)}{f_{X} (\xrm^j; t_1) }}
\approx \mathds{E} \left( \sqrt{
\frac{f_{X} (\xrm; t_2)}{ f_{X} (\xrm; t_1)}}\right) \\
= \int \sqrt{ f_{X} (\xrm; t_1) f_{X} (\xrm; t_2) } d\xrm
= 1-H^2.
\label{eq:mcmc_hellinger}
\end{equation}
From simulations, we find that the number of Monte-Carlo samples $N$ should be at least $10,000$ for convergence. The prediction for the upper bound on stability can then be computed using Eqn.~\ref{eq:smax_normalized}. 
\subsubsection{Bound verification}
\begin{figure}[!h]
\centering
\includegraphics[width=0.8\textwidth]{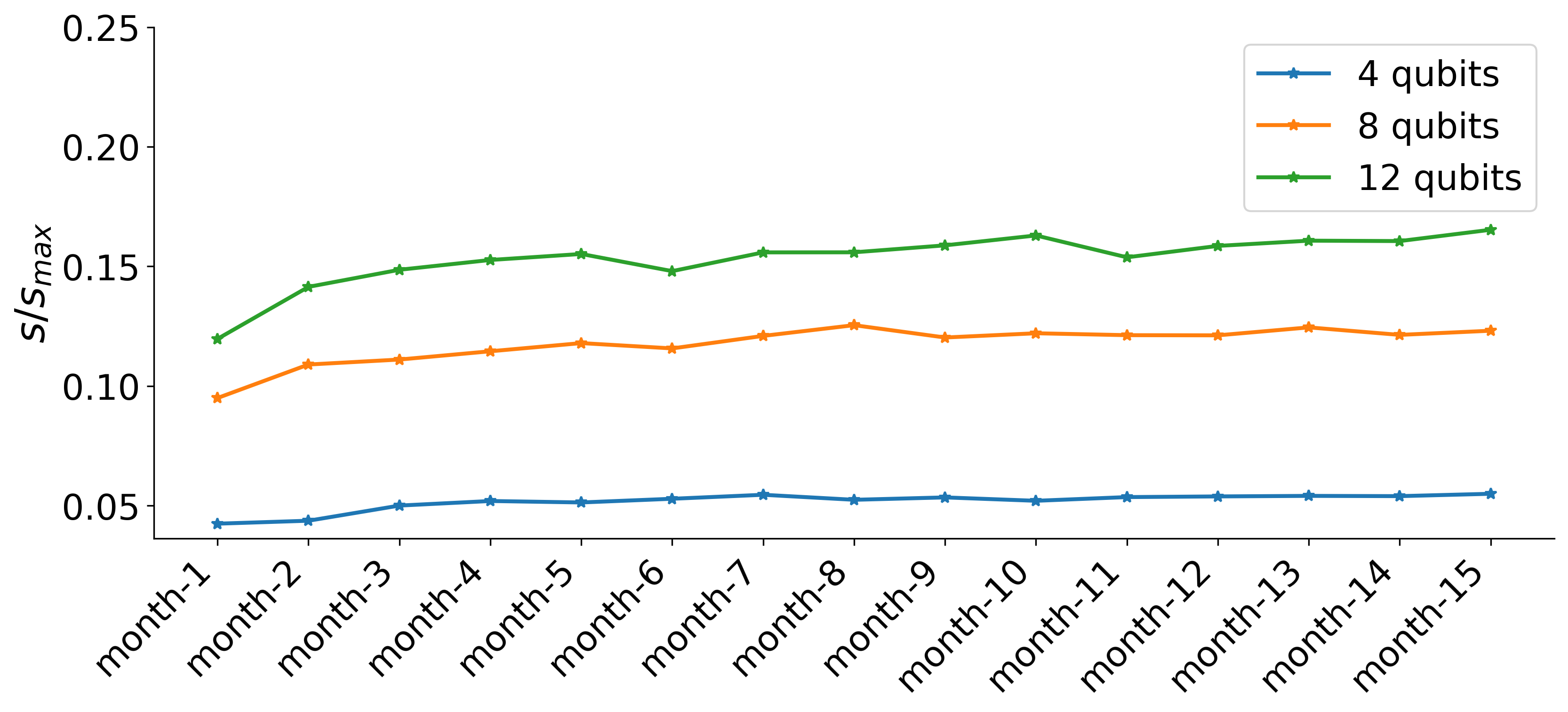}
\caption{Plot of the ratio of $s$ to the predicted upper bound $s_\text{max}$ 
for a circuit simulation with time-varying noise 
(with length of the secret string varying as 4, 8, and 12).
}
\Description{
Plot of the ratio of $s$ to the predicted upper bound $s_\text{max}$.
}
\label{fig:depol_sbysmax}
\end{figure}
Next, we verify the validity of the upper bound $s_\text{max}$ that we computed in the previous section. We simulate the quantum circuit shown in Fig.~\ref{fig:bv_ckt_a} using time-varying noise parameters that are sampled from the joint density $f_{X}(\xrm;t)$, which was modeled using Eqn.~\ref{eq:copulas}. The noisy outcomes obtained from the quantum circuit simulations are analyzed to obtain the time-varying noisy mean of the observable $\braket{\mathcal{O}}_\xrm(t)$, which helps us compute the observed stability metric $s$ using Eqn.~\ref{eq:s_def}. If the ratio $s/s_\text{max}$ stays below 1, then the bound is verified for this numerial simulation exercise.

Fig.~\ref{fig:depol_sbysmax} plots the ratio of the obtained stability metric $s$ to the upper bound $s_\text{max}$ with respect to time. Three different problem sizes, with secret-string length of 4, 8, and 12 respectively, are simulated. In all cases, the results confirm the bound provided by Eqn.~\ref{eq:smax_normalized}. We observe that the ratio is higher for bigger problem size, as $s$ is more sensitive to changes in the Hilbert space dimension than the denominator. For example, in the 4-qubit problem {the ratio ranged from 4\% to 5\%, while for the 8-qubit problem it varied from 9\% to 13\%, and for the 12-qubit problem, it ranged from 12\% to 16\%}.

\subsection{Verification using experimental data}
In this section, we verify the bound on outcome stability using data from a superconducting platform.
\subsubsection{Experimental dataset} 
The experimental noise dataset is derived from the periodically published 
characterization data for IBM's $127$-qubit superconducting device~\cite{krantz2019quantum} called washington~\cite{ibm_quantum_experience_website}. 
It spans {a $16$-month period beginning 1-Jan-2022 and ending 30-Apr-2023}.
The dataset contains the calibration date and time, qubit-wise SPAM error, CNOT gate error, and qubit-wise dephasing times. 
The circuit noise data relevant for our 5-qubit implementation of the \BV~algorithm is shown in Table~\ref{tab:noiseParameters}. 
A sample time-series for one of the noise parameters (SPAM error) is shown in Fig.~\ref{fig:q_3_5_washington_jan2022_apr2023}. 

\begin{figure}[!h]
\centering
\includegraphics[width=0.5\textwidth]{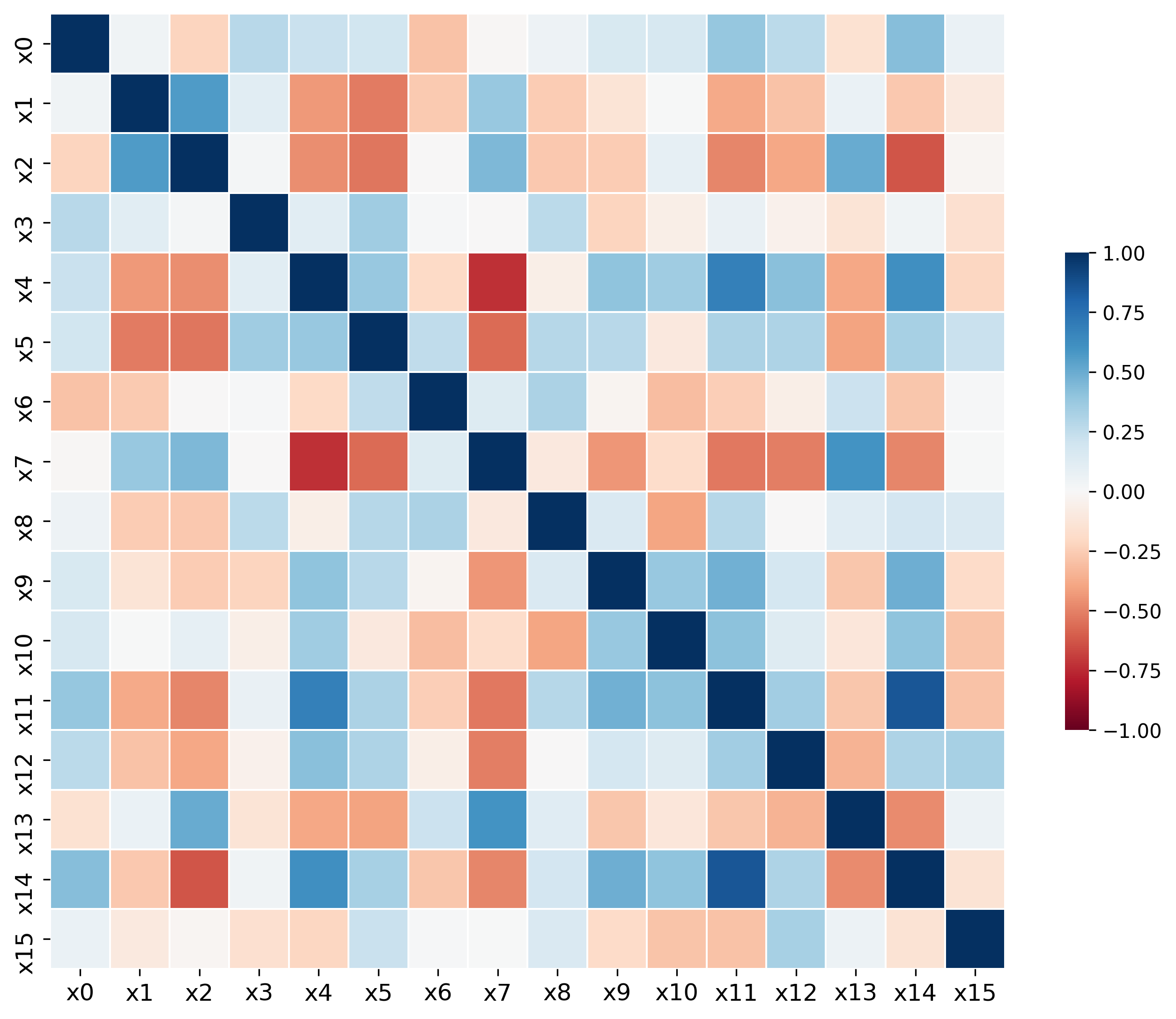}
\caption{Experimental heatmap for correlation between the 16 noise parameters defined in Table~\ref{tab:noiseParameters} as observed in Apr-2023.}
\Description{Heatmap for the Pearson correlation coefficients between the 16 noise parameters}
\label{fig:correlation_matrix}
\end{figure}

The data exhibits significant correlations between them. 
Fig.~\ref{fig:correlation_matrix} shows the correlation structure between the 16 device noise parameters for Apr-2023 (e.g., the error in implementing the Hadamard gate on qubits 0 and 3 were strongly correlated $\approx 86\%$). The error bars on these coefficients is approximately $1/\sqrt{30-1} \approx 0.18$. 
%
\subsubsection{Monte-Carlo sampling}
The experimental dataset described in the previous section is used to empirically estimate the monthly densities for each of the 16 noise parameters using Eqn.~\ref{eq:beta_dist}. 
Alongwith the time-varying correlation structure, these marginal distributions help us estimate the time-varying joint density using Eqn.~\ref{eq:copulas}. 
Although the full distribution cannot be visualized, the significance of the correlation modeling using the copula formalism can be appreciated from Fig.~\ref{fig:dist_with_copula}, which illustrates the significant difference when modeling multi-variate densities with and without correlation. 

We use Monte Carlo simulations (details provided in Sec.~\ref{sec:mcmc_appendix}) to obtain circuit noise samples from this joint density. 
The noise samples generated ($\approx 100,000$ to ensure convergence) 
are used for the estimation of time-varying Hellinger distance.
\subsubsection{Distance estimation}
Next, we compute the Hellinger distance between the time-varying joint densities, over the 16 months spanning {Jan-2022 to Apr-2023}, with Jan-2022 being the reference month. This is used to predict the upper bound $s_\text{max}$ on outcome stability using Eqn.~\ref{eq:smax_normalized}.

\begin{figure}[!hb]
  \centering
\includegraphics[width=0.8\textwidth]{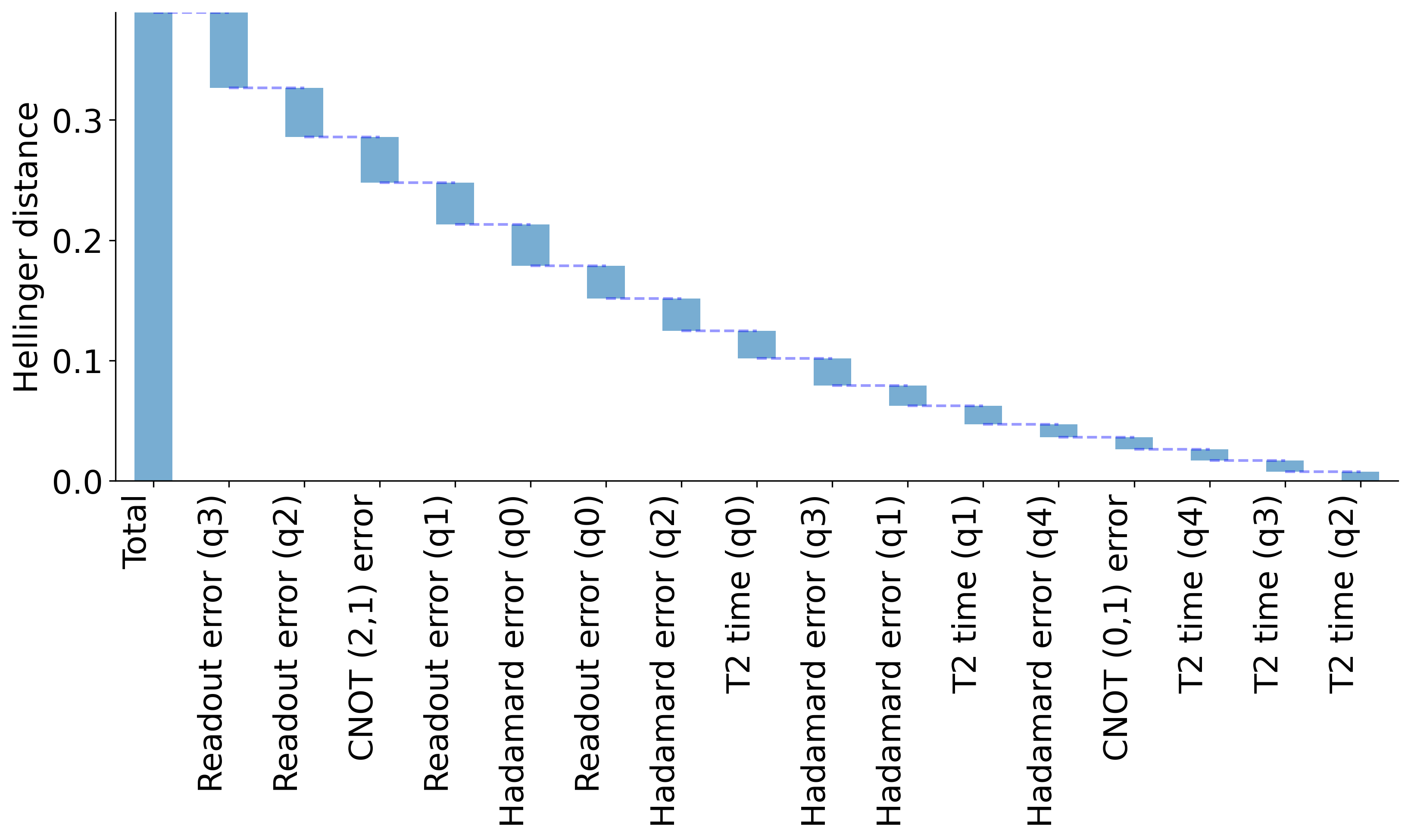}\\
\caption{Contribution made by each noise parameter to the Hellinger distance. The x-axis labels the corresponding noise parameter and the y-axis shows the average Hellinger distance. The computed distance compares the distribution {in Apr-2023 to Jan-2022}. The parameters contribute varying percentages but no single parameter dominates the sum.}
\Description{This empirical plot investigates which metrics for device characterization are responsible for the time variation of a quantum computing platform, rendering it unreliable for physical realization.}
\label{fig:distance_attribution_ind}
\end{figure}

The first 16 columns in Table~\ref{tab:marginal_hellinger19} present the time-varying distance data for univariate (marginal) distributions, while the last three columns contain the same for joint distributions using the method of copulas shown in Eqn.~\ref{eq:copulas}. Fig.~\ref{fig:distance_from_ref} confirms that the curse of dimensionality is most pronounced for $H_r$ (the unmodified Hellinger distance cf. Eqn.~\ref{eq:hellinger_unmodified}) with a range that is only 0.029. 
However, 
$H_a$ (average over the marginals cf. Eqn.~\ref{eq:hellinger_avg}) 
and $H_n$ (normalized distance cf. Eqn.~\ref{eq:hellinger_normalized}) 
do not suffer from this (the range is 0.51 for $H_n$ and 0.20 for $H_a$). 
Fig.~\ref{fig:distance_attribution_ind} further breaks down the contributions from the 16 noise sources. It can be seen that the measurement operation (quantified by SPAM) accounts for the largest contribution to device unreliability.
\subsubsection{Noise model}
The main sources of noise when implementing the \BV~circuit as shown in Fig.~\ref{fig:bv_ckt_b} on superconducting hardware are: 
qubit decoherence, 
noise during the implementation of single-qubit gates like Hadamard gate, 
noise during the implementation of entangling gates like CNOT gate, 
noise associated with state preparation and measurement (SPAM), 
and noise during the implementation of the Z gate. 
Among these, the noise from state preparation and measurement (SPAM) typically contributes most significantly to errors~\cite{bravyi2021mitigating}. 
The second largest source of error is the imperfect implementation of entangling gates, such as the CNOT gate. 
Following these, the impact of noise decreases significantly, by two orders of magnitude or more, for single-qubit rotations and qubit decoherence when the qubits are idle~\cite{van2023probabilistic, bravyi2021mitigating}. 
The $\mathds{Z}$ gate is a software-based operation~\cite{mckay2017efficient} and error-free. 
We will next describe each of these types of noise.

\textit{(i) Qubit decoherence}: Decoherence in quantum systems refers to the loss of unitarity during state evolution, often quantified using two metrics: amplitude relaxation time ($T_1$), and de-phasing time ($T_2$). $T_1$ measures the amplitude attenuation, indicating the probability that a qubit will transition from an excited state to the ground state over time, modeled by an exponential decay function. $T_2$ tracks the time it takes for the superposition state of a qubit to decay, affecting the off-diagonal elements of the density matrix. In practice, $T_2$ dominates and is modeled using Kraus operators as~\cite{nielsen2002quantum}: $\mathcal{E}(\rho) = E_0 \rho E_0^{\dagger} + E_1 \rho E_1^{\dagger}$ 
where 
$E_0 = 
\begin{pmatrix}
1 & 0\\
0 & \exp(-t/2T_2)
\end{pmatrix}E_1 = 
\begin{pmatrix}
0 & 0\\
0 & \sqrt{1-\exp(-t/T_2)}
\end{pmatrix}$, and $t$ is the time scale of the circuit-idle time for a qubit under superposition. 
There are five qubits in our circuit whose de-phasing time is given by $\xrm_{6}, \cdots, \xrm_{10}$ in Table~\ref{tab:noiseParameters}.

\textit{(ii) Single-qubit gate}: Gate fidelity $F_G$ measures the accuracy with which a quantum operation transforms the register state. Many methods \cite{chuang1997prescription, emerson2005scalable, knill2008randomized, gambetta2012characterization, mckay2019three} have been developed to characterize noise in gate operations. 
Randomized benchmarking is widely used to measure the quantum state survival probability following a sequence of randomly selected Clifford elements by fitting the resulting sequence of fidelity measurements to a linear model that eliminates the influence of state preparation and measurement errors \cite{magesan2011scalable}. These parameters are then used to estimate the average error per Clifford gate with more sophisticated methods, such as interleaved randomized benchmarking, capable of estimating the error of a specific gate. The resulting gate fidelity $F_G$ is defined by the error per Clifford gate $\epsilon_G$:
\begin{equation}
F_G = 1 - \epsilon_{G}.
\label{eqn:fg}
\end{equation}
In the context of the \BV~circuit, the $\mathds{Z}$-gate is noiseless as discussed before. The only single-qubit noisy gate is the Hadamard gate ($\mathds{H}$). The impact of noise on a single qubit density matrix $\rho$ is modeled using a depolarizing channel $\mathcal{E}(\rho) = F_G \mathds{H}\rho\mathds{H}^\dagger + (1-F_G)\mathds{I}/2$
where $\mathds{I}$ is the identity operation. The matrix representation for the ideal Hadamard operation is given by:
$\mathds{H} = \frac{1}{\sqrt{2}}
\begin{pmatrix}
1&1\\
1&-1\\
\end{pmatrix}$.
There are five Hadamard gates in our circuit whose gate fidelity ($F_G$) are given by $\xrm_{11}, \cdots, \xrm_{15}$ in Table~\ref{tab:noiseParameters}.

\textit{(iii) CNOT gate}: 
%
The noise in the CNOT gate is modeled using a symmetric 2-qubit depolarizing model:
\begin{equation}
\begin{split}
\mathcal{E}_\text{CNOT}(\cdot) =& F_G^2 (\cdot) +
\frac{F_G(1-F_G)}{3}\left[
\sum \limits_{\mathds{P}_T}  (\mathds{I}_C \otimes \mathds{P}_T) (\cdot) (\mathds{I}_C \otimes \mathds{P}_T) + 
\sum \limits_{\mathds{P}_C}  (\mathds{P}_C \otimes I_T ) (\cdot) (\mathds{P}_C \otimes I_T )
\right]\\
&+\frac{ (1-F_G)(1-F_G) }{9} \sum \limits_{\mathds{P}_C, \mathds{P}_T}  (\mathds{P}_C \otimes \mathds{P}_T) (\cdot) (\mathds{P}_C \otimes \mathds{P}_T)\\
\end{split}
\end{equation}
In this expression, we use $\mathds{P}_C, \mathds{P}_T \in \{\mathds{X}, \mathds{Y}, \mathds{Z}\}$ as the single-qubit Pauli operators excluding identity $\mathds{I}$, acting on the control (C) and target (T) qubits respectively. There are two CNOT gates in our circuit whose fidelity ($F_G$) is given by $\xrm_4$ and $\xrm_5$ in Table~\ref{tab:noiseParameters}.

\textit{(iv) SPAM  fidelity:} State preparation and measurement fidelity $F_\text{SPAM}$ quantifies the accuracy with which a target quantum state is prepared and measured in the register and this can be quantified by the quantum state fidelity. Tomographic methods are informative for measuring the fidelity of general quantum states but these approaches typically scale steeply with the size of the register \cite{aaronson2019shadow}. An alternative method measures the fidelity of a more restrictive set of states, which in the extreme limit, may be reduced to characterizing a fiducial computational basis state, such as $|0\rangle^{\otimes n}$. A simple and highly efficient characterization defines the state preparation and measurement (SPAM)  fidelity $F_\text{SPAM}$ as the probability to measure the $n$-bit string matching the fiducial state and the error $\epsilon_\text{SPAM}$ as the probability of observing any other outcome:
\begin{equation}
F_\text{SPAM} = 1 - \epsilon_\text{SPAM}.
\end{equation}
Notably, readout error is implicitly included in this characterization method. 
SPAM noise can be modeled using channel formalism~\cite{smith2021qubit}. 
The noise channel is an effective model that assumes the output density matrix post circuit execution is corrupted by qubit-specific Kraus operators that are characterized by the qubit-specific SPAM fidelity. Specifically, qubit $q$ is impacted by two Kraus operators $M_0$ and $M_1$:
\begin{equation}
\begin{split}
M_0 =& \sqrt{ F_\text{SPAM}}\ket{0}\bra{0}+\sqrt{1-F_\text{SPAM}}\ket{1}\bra{1}\\
M_1 =& \sqrt{1-F_\text{SPAM}}\ket{0}\bra{0}+\sqrt{F_\text{SPAM}}\ket{1}\bra{1}\\
\end{split}
\end{equation}
where $F_\text{SPAM}$ represents the SPAM fidelity of qubit $q$. The probability of measuring 0 is given by $\text{Tr}\left(M_0^\dagger M_0 \rho\right)$ while the probability of measuring 1 is given by $\text{Tr}\left(M_1^\dagger M_1 \rho\right)$. Neglecting inter-qubit cross-talk, we then have the $n$-qubit SPAM noise channel representation (assuming separability):
\begin{equation}
\mathcal{E}(\cdot) = \left[\bigotimes\limits_{q=0}^{n} \mathcal{E}_q^\text{SPAM}\right](\cdot)
\end{equation}
where $\mathcal{E}_q^\text{SPAM}(\cdot) = M_0 (\cdot) M_0^\dagger + M_1 (\cdot) M_1^\dagger$ denotes a single-qubit SPAM noise channel for qubit $q$. The SPAM fidelity ($F_G$) for the 4 qubits in our circuit which get measured (the ancilla qubit does not get measured) are given by $\xrm_{0}, \cdots, \xrm_{3}$ in Table~\ref{tab:noiseParameters}.

\subsubsection{Circuit simulation}
\begin{figure}
\centering
\includegraphics[width=0.5\textwidth]{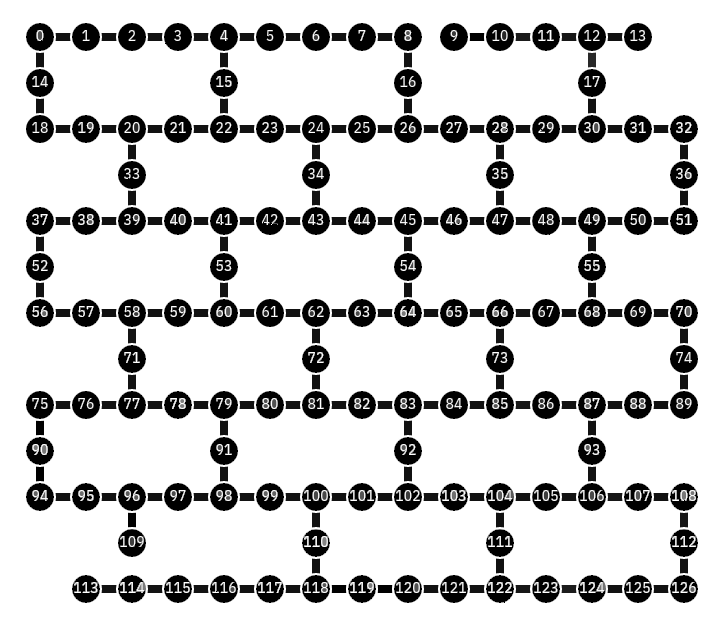}
\caption{Schematic layout of the $127$-qubit \textit{washington} device produced by IBM. Circles denote register elements and edges denote connectivity of 2-qubit operations. The register elements $0$, $1$, $2$, $3$, and $4$ in Fig.~\ref{fig:bv_ckt_b} are mapped to the physical qubits $4$, $3$, $2$, $1$, and $0$, respectively, in the diagram above. The CNOT gates used in the circuit connect the physical qubits $(0,1)$ and $(2,1)$, in the diagram above, where the first number represents the control qubit and the second one represents the target qubit.}
\Description{Schematic layout of the 127-qubit washington device produced by IBM.}
\label{fig:wash_bw}
\end{figure}

Next, we obtain statistical samples for the outcome stability $s$ to verify if the bound is satisfied. 
To achieve this, we repeatedly execute the quantum circuit shown in Fig.~\ref{fig:bv_ckt_b} in Qiskit~\cite{aleksandrowicz2019qiskit} using the noise samples generated through the Monte Carlo procedure for the noise model we described in the previous section using Eqn.~\ref{eq:copulas}. 
For simulation purposes, the logical qubits $0$, $1$, $2$, $3$, and $4$ are mapped to the physical qubits $4$, $3$, $2$, $1$, and $0$ in Fig.~\ref{fig:wash_bw}. 
The two CNOT gates shown in Fig.~\ref{fig:bv_ckt_b} connect the physical qubits $(0,1)$ and $(2,1)$, where the first number represents the control qubit and the second one represents the target qubit. 
The results obtained are used to calculated the time-varying mean of the observable $\braket{O}_\xrm(t)$ and thence, the time-varying stability metric $s(t_1, t_2) = |\braket{O}_\xrm(t_1)-\braket{O}_\xrm(t_2)|$. 

\subsubsection{Bound verification}\label{sec:unreliable}
\begin{figure}[!h]
\centering
\includegraphics[width=0.9\textwidth]{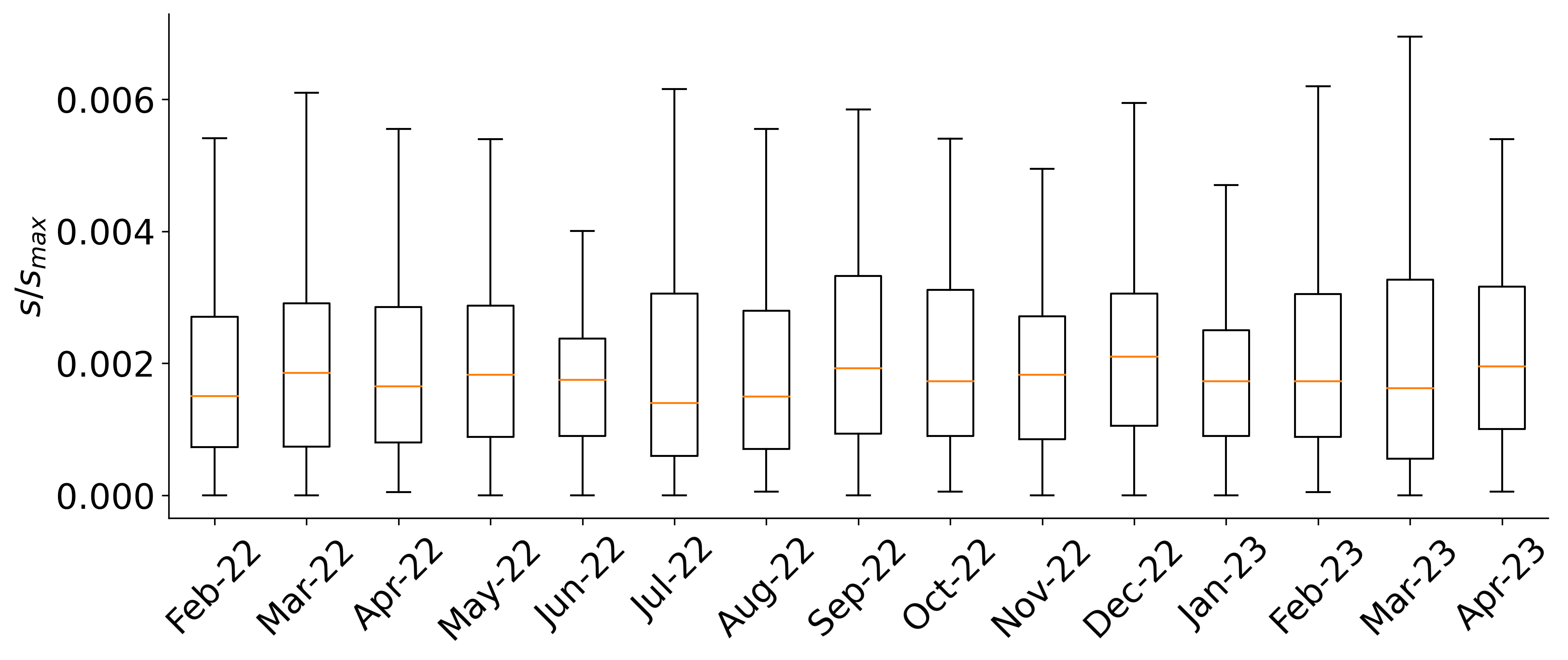}
\caption{ Simulations of the ratio of the stability metric $s$ to the predicted upper bound $s_{max}$ for a 4-qubit \BV~circuit using the noise characterization data from the IBM washington platform {from 1-Jan-2022 to 30-Apr-2023}. Ratio values below unity confirm the predicted upper bound. 
}
\Description{We implemented a 4-qubit \BV~circuit on the IBM washington platform and modeled the monthly joint noise distribution (across the 16 circuit parameters).}
\label{fig:s_by_smax}
\end{figure}

Fig.~\ref{fig:s_by_smax} shows the box-and-whisker plot for the ratio of $s$ to $s_{\text{max}}$ for each month. 
The central box at each point signifies the inter-quartile range (IQR), with its lower and upper edges representing the first (Q1) and third quartiles (Q3), respectively. The median is indicated by a line within the box. All the observed values remain well below unity, thereby verifying the bound. 

The ratio of approximately 0.007 indicates that the predicted bound is significantly higher than necessary. Applying this stringent bound in manufacturing could result in over-engineering. However, using the bound offers clear advantages. Without this bound, it is difficult to estimate the device improvement needed for stable outcomes. The only other way is to conduct millions of experiments across a very long time which is not feasible. If the device stays within our predicted bound, program outcome will be stable with high statistical confidence.

There are two reasons why the upper bound is not reached in our experiment. Firstly, the noise density charaterization has been done at monthly frequency, using daily data from IBM. At monthly time-scale, the variance does not simply increase with time (unlike Fig.~\ref{fig:depol_sbysmax}) but exhibits randomness. This impacts Eqn.~\ref{eq:abs_integrand} in terms of the tightness of the bound. 
Secondly, as might be expected, each application of Holder's inequality introduces additional loss of tightness. 

\begin{table}[htp]
\centering
\begin{tabular}{|p{1.4cm}|p{6.4cm}|}
    \hline 
\textit{Parameter} & \textit{Description} \\ \hline
$\xrm_{0}, \cdots, \xrm_3$ & SPAM  fidelity for register element 0, 1, 2, 3 \\ \hline
$\xrm_{4}, \xrm_{5}$ & CNOT gate fidelity for (control 0, target 1) and  (control 2, target 1) \\ \hline
$\xrm_{6}, \cdots, \xrm_{10}$ & dephasing time for register element 0,1,2,3,4\\ \hline
$\xrm_{11}, \cdots, \xrm_{15}$ & Hadamard gate fidelity for register element 0,1,2,3,4  \\ \hline
\end{tabular}
\medskip\medskip\medskip
\caption{Circuit noise parameters}
\Description{Circuit noise parameters}
\label{tab:noiseParameters}
\end{table}

\begin{figure}[!h]
\centering
\includegraphics[width=0.8\textwidth]{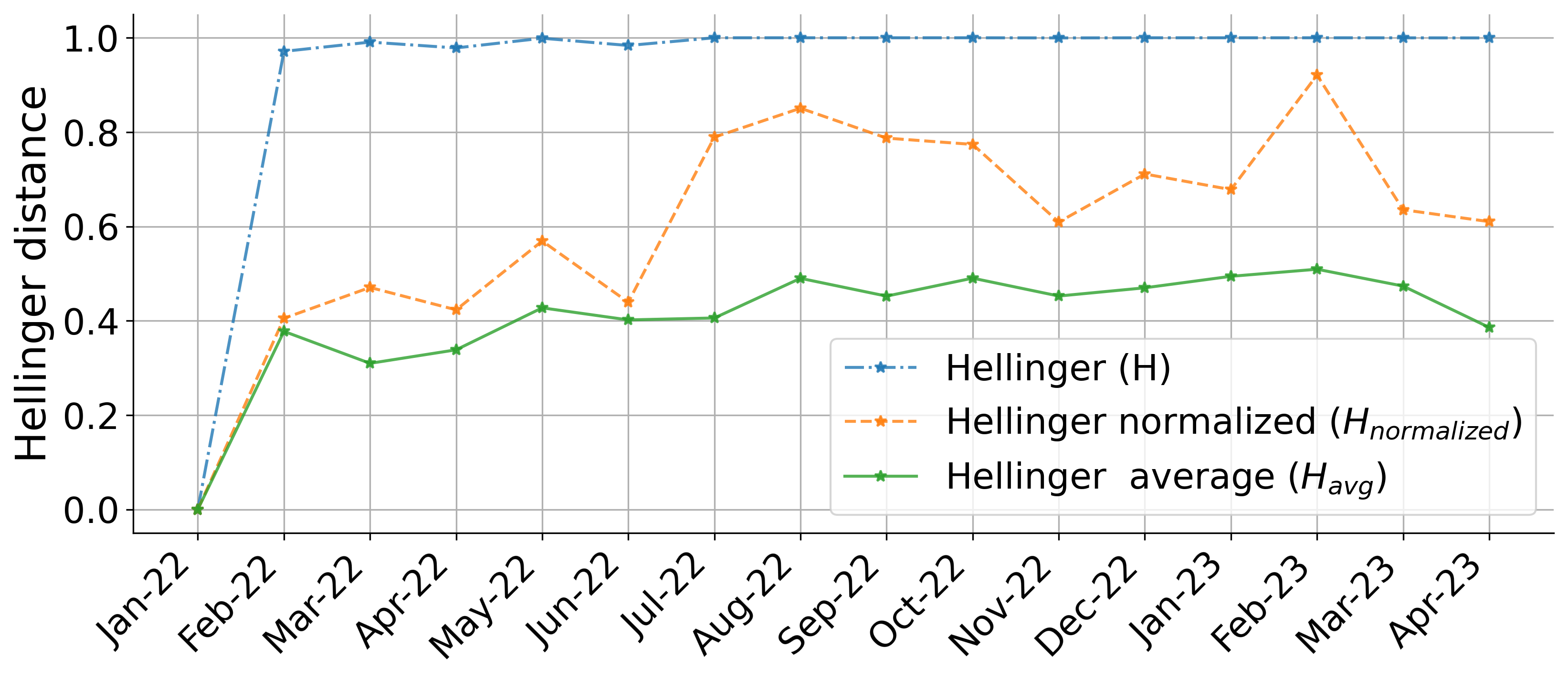}
\caption{
Plots of the three Hellinger distance measures, with the unmodified measure $H$ (blue line) being insensitive due to dimensionality issues. The blue and orange lines both capture the correlation structure of the joint distribution (with the orange line being normalized to enhance discrimination power), while the green line lacks correlation capture. 
}
\Description{The figure presents results of reliability testing on a 127-qubit superconducting platform}
\label{fig:distance_from_ref}
\end{figure}

%

\section{Conclusion}\label{sec:conc}
Current experimental demonstrations often rely on quantum circuits calibrated immediately prior to program execution or tuned during run-time. However, the resulting circuits and device calibrations are implicitly dependent on the noise parameters, which fluctuate significantly over time, across the register, and with varying technology. In this study, we derived and validated bounds on outcome stability using the \BV~algorithm as a test circuit. We used noise data from a superconducting device and showed that non-stationary fluctuations in noise parameters are a significant concern for stable quantum computing. Such unreliable devices are not suitable for reproducible error attribution or uncertainty quantification. 

We verified our method using IBM's 127-qubit washington device only and did not compare across devices. 
Although comparing reliability across devices is a worthy goal, our paper focuses on methods for quantifying reliability and assessing the impact of device unreliability on computational stability. 
Better-engineered devices than washington may exist, although superconducting devices share similar noise profiles and remain far from delivering accurate computations. 
Our results are expected to be applicable to other devices because, firstly, well-characterized devices that share similar noise profiles exhibit comparable accuracy and stability in outcomes. 
Secondly, the theory is not specific to any one device; the choice of washington as a verification platform  does not influence the theory presented in Section 2 or Appendix A. 
However, the methodology does require well-characterized device noise data as input. Also, we used a benchmark circuit, namely the \BV~circuit for our study. A comprehensive reliability assessment that is applicable for all possible applications was beyond our scope.  

An important aspect to consider is the adequacy of the noise model used, even when dealing with a specific benchmark circuit. 
It is crucial to acknowledge that achieving a 100\% complete characterization of noise is difficult. For instance, incorporating the effects of cross-talk between non-utilized and distant qubits does not necessarily enhance predictive accuracy. A better focus is sufficiency and prioritization of noise sources. It is essential to ensure that the primary noise sources affecting the circuit are thoroughly characterized. The strategy we used in this paper was to focus on the major noise contributors that account for most of the variability in the outcomes. Note that failing to account for major sources of noise can result in violations of the bounds predicted in this study. For instance, if we overlook SPAM (state preparation and measurement) errors and focus solely on single qubit rotation errors, the predicted upper bound would approach zero, which would consistently fail verification. 

We defined a device to be $\varepsilon$-reliable when $H < \varepsilon$. The threshold $\varepsilon$ depends on the tolerance for instability. {For example, if we require that the mean of the observable of the \BV~circuit at time $t_2$ be within $\pm 0.25$ of the output at time $t_1$, then the distribution similarity measure $H_n(t_1, t_2)$ (normalized measure) needs to be less than 0.022. 
However, from Fig.~\ref{fig:distance_from_ref}, $H_n$ varied between $0.41$ and $0.92$ over the one-and-a-half year period Jan-2022 to Apr-2023}, and this indicates that the device was not reliable for consistently reproducing the  statistical mean in the context of the \BV~benchmark.

Lastly, we note that this study emphasized phenomenological modeling to analyze reliability and stability. 
The methods developed are equally applicable to logical qubits, which comprise thousands of physical qubits, as long as a consistent, modular approach to noise characterization is undertaken. 
This mirrors approaches in conventional computer engineering, which analyze chip-level fault rates instead of semi-conductor level modeling. 
The modular techniques presented in this paper will remain relevant as we progress into the fault-tolerant era.

%


\begin{acks}
This material is based upon work supported by the US Department of Energy, Office of Science, National Quantum Information Science Research Centers, Quantum Science Center.
\end{acks}

\bibliographystyle{unsrt}
\bibliography{references.bib}

\appendix
\section{Stability bounds}\label{sec:upperbound}
Let $s(t_1, t_2)$ be the absolute difference in the mean of the observable obtained from the execution of the noisy quantum program at times $t_1$ and $t_2$. Then,
\begin{equation}
\begin{aligned}
s^2(t_1, t_2) &= \left( \braket{O}_{t_1} - \braket{O}_{t_2} \right)^2\\
&= \left( \int \braket{O_{ \xrm}} f_X(\xrm; t_1)\text{dx} - \int \braket{O_{ \xrm}} f_X(\xrm; t_2)\text{dx} \right)^2\\
&\leq \left( \int \left| \braket{O_{ \xrm}}  \{ f_X(\xrm; t_1)\text{dx} - f_X(\xrm; t_2)\} \right| \text{dx} \right)^2.
\end{aligned}
\label{eq:abs_integrand}
\end{equation}
In the last step, the inequality stems from the absolute value on the integrand. Now, per H\"older's inequality, if $m, n \in [1, \infty)$ and \[\frac{1}{m}+\frac{1}{n}=1,\] then:
\[ \int \left| f(x) g(x) \right|dx  \leq  \left( \int |f(x)|^m dx\right)^{1/m} \left( \int |g(x)|^n dx\right)^{1/n}.\]
Thus, our inequality becomes:
\begin{equation}
\left( \int \left| \braket{O_{ \xrm}}  \{ f_X(\xrm; t_1)\text{dx} - f_X(\xrm; t_2)\} \right| \text{dx} \right)^2 \leq \left[ \left( \int  | \braket{O_{ \xrm}} |^m \text{dx}\right)^{1/m} \left( \int | f_X(\xrm; t_1) - f_X(\xrm; t_2) |^n \text{dx}\right)^{1/n} \right]^2.
\end{equation}
Now, let $m \rightarrow \infty, n=1$ and define:
\begin{equation}
c= \underset{\xrm}{\text{sup}} |\braket{O_{ \xrm}}|.
\end{equation}
Clearly, 
\begin{equation}
\lim\limits_{m \rightarrow \infty} \left( \int |\braket{O_{ \xrm}}|^m\text{dx}\right)^{1/m}\leq \lim\limits_{m \rightarrow \infty} \left( \int c^m\text{dx}\right)^{1/m} = c \left( \lim\limits_{m \rightarrow \infty}  \left( \int \text{dx}\right)^{1/m} \right)= c.
\label{eq:c_factor}
\end{equation}
Thus, we have
\begin{equation}
\begin{aligned}
s(t_1, t_2)^2 
&\leq  \lim\limits_{m\rightarrow \infty, n \rightarrow 1} \left( \left( \int  | \braket{O_{ \xrm}} |^m \text{dx}\right)^{1/m} \left( \int | \{ f_X(\xrm; t_1)\text{dx} - f_X(\xrm; t_2)\} |^n \text{dx}\right)^{1/n} \right)^2\\
&\leq  \left( 
\lim\limits_{m\rightarrow \infty}
\left( \int  | \braket{O_{ \xrm}} |^m \text{dx}\right)^{1/m} 
\left( 
\lim\limits_{n \rightarrow 1}
\int | \{ f_X(\xrm; t_1)\text{dx} - f_X(\xrm; t_2)\} |^n \text{dx}\right)^{1/n} \right)^2\\
&= c^2 \left( \int | f_X(\xrm; t_1) - f_X(\xrm; t_2)| \text{dx} \right)^2 \\
&=c^2 \left( \int \left| \sqrt{f_X(\xrm; t_1)} - \sqrt{f_X(\xrm; t_2)} \right| \left( \sqrt{f_X(\xrm; t_1)} + \sqrt{f_X(\xrm; t_2)} \right) \text{dx} \right)^2\\
&=c^2 \int \left( \sqrt{f_X(\xrm; t_1)} - \sqrt{f_X(\xrm; t_2)} \right)^2 \text{dx}\int \left( \sqrt{f_X(\xrm; t_1)} + \sqrt{f_X(\xrm; t_2)} \right)^2 \text{dx} \;\;\;\;\text{ (applying H\"older's inequality with m=n=2)} \\
&=c^2\int \left( f_X(\xrm; t_1)+ f_X(\xrm; t_2) - 2\sqrt{f_X(\xrm; t_1)}\sqrt{f_X(\xrm; t_1)} \right) \text{dx} \int \left( f_X(\xrm; t_1)+ f_X(\xrm; t_2) + 2\sqrt{f_X(\xrm; t_1)}\sqrt{f_X(\xrm; t_1)} \right) \text{dx}\\
&=c^2 \left( 1+ 1 - 2\int \sqrt{f_X(\xrm; t_1)}\sqrt{f_X(\xrm; t_1)}\text{dx} \right) \left( 1+ 1 + 2\int \sqrt{f_X(\xrm; t_1)}\sqrt{f_X(\xrm; t_1)}\text{dx} \right)\\
&=4c^2H_{X}^2(2-H_{X}^2),
\end{aligned}
\end{equation}
where, for notational clarity, we use:
\[H_X^2 = H^2_X(t_1, t_2) = 1-\int \sqrt{f_X(\xrm; t_1)}\sqrt{f_X(\xrm; t_2)}dx\]
Equation (\ref{eq:smax_normalized}) upper bounds temporal variations in the expected mean of the circuit outcome with respect to the statistical reliability of the device noise parameters. As expected, the temporal variations vanish when $H_X = 0$ and become maximal when $H_X = 1$. The bound itself is monotonic with the greatest change occurring for the greatest distance. Note that $c=\underset{\xrm}{\text{sup}} |\braket{O_{ \xrm}}|$ has no dependence on the time-varying distribution $f_X(\xrm; t)$ but it does depend on the noise parametrization for the circuit. Thus the observable stability $s$ is always upper bounded by 
\[s_{\text{max}} = 2c H_{X} \sqrt{2-H_{X}^2},\] 
an upper bound determined by the degree of time-variation of the noise parameters.

\begin{figure*}[!h]
\centering
\includegraphics[width=.5\textwidth]{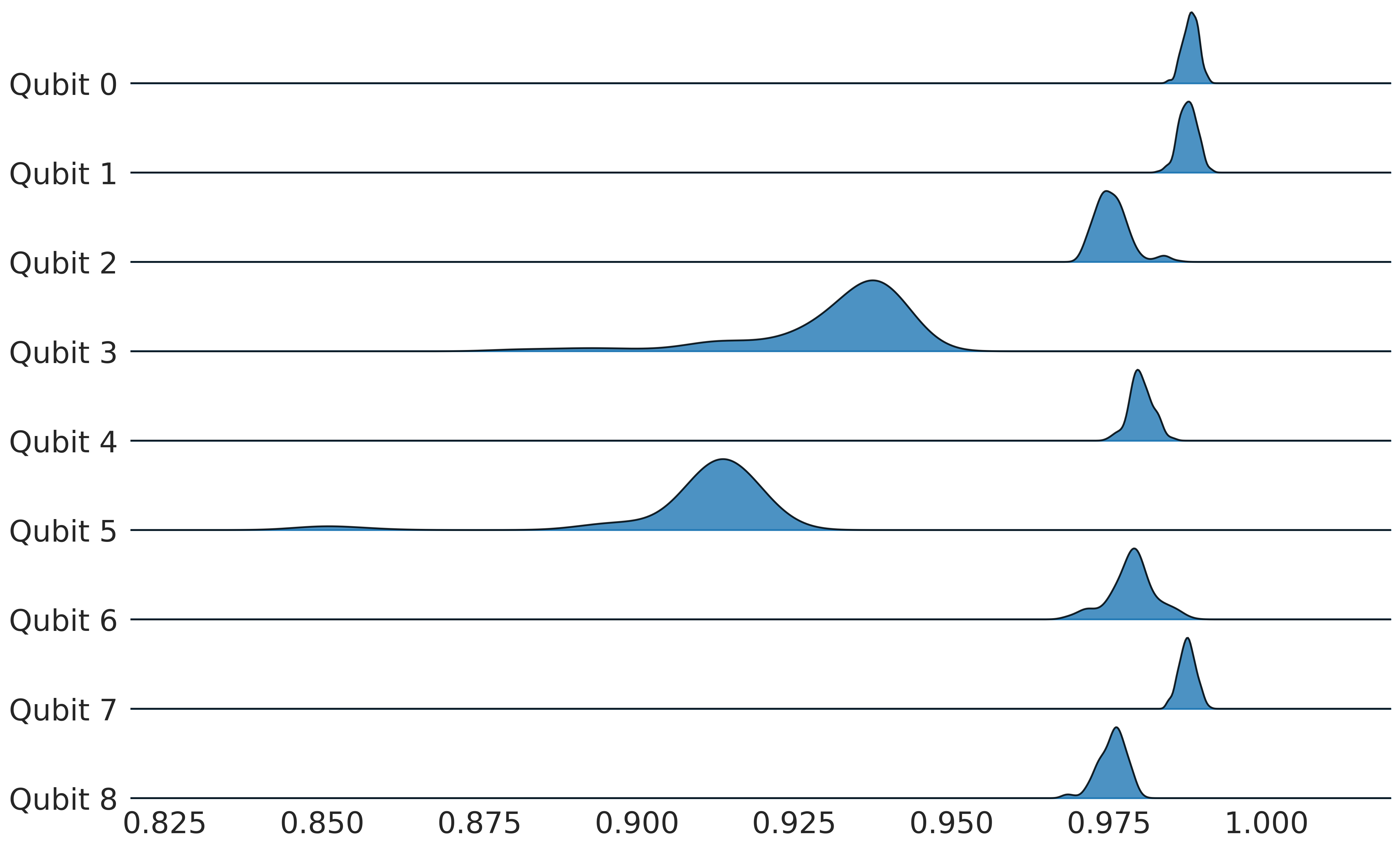}
\medskip
\caption{Unimodal, skewed SPAM fidelity distributions for a superconducting device (IBM toronto) for qubits $0-8$ as measured on 8 April 8 2021, between 8:00-10:00pm (UTC-05:00).}
\Description{Unimodal, skewed SPAM fidelity distributions for a superconducting device (IBM toronto for qubits $0-8$ as measured on 8 April 8 2021, between 8:00-10:00pm (UTC-05:00).}
\label{fig:f0f1_toronto_qubit_0_onwards_spruce_2021}
\end{figure*}
\vspace{0.5in}
\begin{figure}[!h]
\centering
\includegraphics[width=0.5\textwidth]{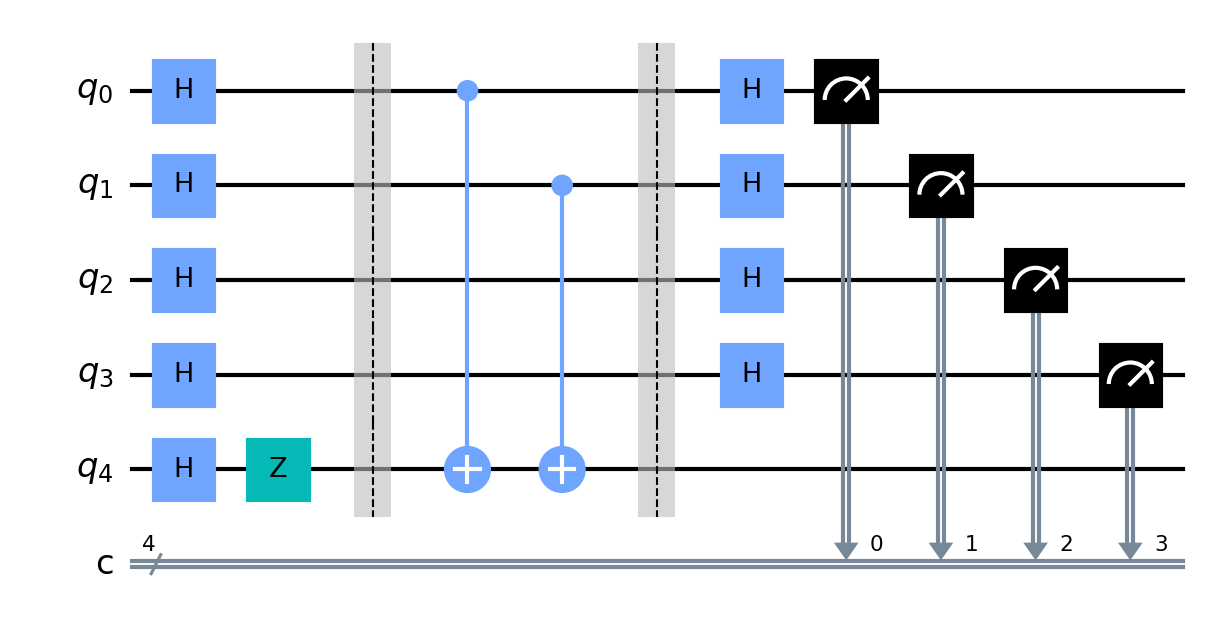}
\caption{
Schematic of the \BV~circuit compiled and simulated using IBM qiskit with time-varying circuit noise reflecting the superconducting washington device.
}
\Description{
Quantum circuit for the implementation of the \BV~algorithm.
}
\label{fig:bv_ckt_b}
\end{figure}
\vspace{0.5in}
\section{Time-invariance for wide-sense stationary noise processes}\label{sec:WSS}
For the \BV~problem, $\braket{O}_t$ is time-invariant if the noise is independent and wide-sense stationary (WSS) and the quantum channel, as a function of the noise parameter $\xrm$, is a first-order polynomial (e.g. depolarizing channel). This is because:
\begin{equation}
\begin{split}
\braket{O}_t 
=& \int \braket{O_{X}} f_{X}(\xrm;t) d\xrm\\
=& \prod_{i=1}^n \int \bra{r_i} \mathcal{E}_{\xrm_{i}} (\ket{r_i}\bra{r_i} )\ket{r_i} f_{X_i}(\xrm_{i};t) d\xrm_{i}\\
=& \prod_{i=1}^n \int \text{(A first-order polynomial in $x_i$)} f_{X_i}(\xrm_{i};t) d\xrm_{i},
\end{split}
\end{equation}
which is time-invariant because of the wide-sense stationarity assumption, which states that WSS (wide-sense stationary) random processes possess a constant first-order moment and auto-covariance. Thus,
\begin{equation}
s = |\braket{O}_t - \braket{O}_0| = 0 \;\;\;\;\forall t.
\end{equation}
%
\section{Curse of dimensionality}\label{sec:dimcurse}
Consider $d$ independent and identically distributed circuit error parameters $\{\xrm_1, \cdots, \xrm_d\}$, whose marginal (uni-variate) distributions are given by $f_{X_i}(\xrm;t)$. Let $h$ be the Hellinger distance between the marginals at time $t_1$ and $t_2$. Thus, \[H_{X_i}(t_1, t_2)=h \;\;\;\; \forall i\]
Since the parameters are independent:
\begin{equation}
\begin{split}
\log \left( 1 - H_{X}^2 \right) &= -d |\log (1-h^2)|\\
\Rightarrow H_{X} &= \sqrt{1-\exp\left[ -d |\log (1-h^2)|\right]}.
\end{split}
\label{eq:dimcurse}
\end{equation}
Thus the distance approach $1$ quickly as the number of dimensions increases. 

\section{Hyper-parameters}\label{sec:beta_hyper_params}
If the initial mean $\mu_0$ and variance $\sigma_0^2$ is known, then a beta distribution with parameters $\alpha_0$ and $\beta_0$ requires:
\begin{equation}
\begin{split}
\mu_0 =& \frac{\alpha_0}{\alpha_0+\beta_0}\\
\sigma^2_0 =& \frac{\alpha_0 \beta_0}{ (\alpha_0+\beta_0)^2 (1+\alpha_0+\beta_0)}\\
\end{split}
\end{equation}
which yields:
\begin{equation}
\begin{split}
\alpha_0 =& \mu_0 \left[ \frac{\mu_0(1-\mu_0)}{\sigma_0^2} -1\right]\\
\beta_0 =& (1-\mu_0) \left[ \frac{\mu_0(1-\mu_0)}{\sigma_0^2} -1\right]\\
\end{split}
\end{equation}
Further, if $\omega$ is the multiple by which the variance increases ($\sigma_T^2 = \omega \sigma_0^2$) in between two consecutive calibrations (e.g. they could be spaced 24 hours apart), then:
\begin{equation}
\begin{split}
\omega \sigma_0^2 =& \frac{\alpha_T \beta_T}{ (\alpha_T+\beta_T)^2(1+\alpha_T+\beta_T)}\\
=& \frac{\mu_0(1-\mu_0)}{1+\alpha_T+\beta_T}\\
=& \frac{\mu_0(1-\mu_0)}{1+ \frac{\alpha_0+\beta_0}{k_0+T} }\\
\Rightarrow k_0 =& \frac{\phi-1}{\phi/\omega-1}-T\\
\end{split}
\end{equation}
where $\phi = \frac{\mu_0(1-\mu_0)}{\sigma_0^2}$.

Thus, we have determined the time-variation of the parameters characterizing the beta distribtion:
\begin{equation}
\begin{split}
\alpha_t =& \frac{\alpha_0}{k_0 + t}\\
\beta_t =& \frac{\beta_0}{k_0 + t}\\
\end{split}
\end{equation}
from which the time-variation of the mean and variance follow.

\section{Exponential deterioration in stability on unreliable devices}\label{sec:scalability}
We defined outcome stability $s$ as:
$$
s = \left| \braket{\mathcal{O}}_\xrm (t_1) - \braket{\mathcal{O}}_\xrm (t_2) \right|
$$
where $\braket{\mathcal{O}}_\xrm$ is the mean of the quantum observable 
and 
$\braket{\mathcal{O}}_\xrm (t)$ is its average over the noisy density at time $t$.

For the \BV~problem, we have shown previously in Eqn.~\ref{eq:bv_ox_def} that:
$
\braket{\mathcal{O}}_\xrm = \prod\limits_{i=1}^{n} \left(1-\frac{\xrm_i}{2}\right)
$ 
where the de-polarizing parameter for qubit $i$ is denoted by $\xrm_i$, $n$ is the register size, and, 
$
\braket{\mathcal{O}}_\xrm (t) = \int\limits_\xrm \prod\limits_{i=1}^{n} \left(1-\frac{\xrm_i}{2}\right)
f(\xrm_1, \cdots, \xrm_n; t).
$ Using the transformation $ y_i = (\xrm_i - \mu_i)/\sigma_i$ 
(where $\mu_i$ and $\sigma_i$ denote the mean and standard-deviation, respectively, of the de-polarizing noise parameter $\xrm_i$), 
we can re-write the time-varying mean of the observable for the $n$-qubit \BV~problem as:
\begin{equation}
\begin{split}
\braket{\mathcal{O}}_\xrm(t) =& \mathds{E}\left[ \prod\limits_{i=1}^n \left( 1 - \frac{\xrm_i}{2}\right)\right] = \sum\limits_{k=1}^{n+1} \mathds{E}(\chi_k)\\
\end{split}
\end{equation}
where
\begin{equation}
\chi_k = (-1)^{k-1} 
\sum\limits_{1 \leq i_1 < i_2 \cdots < i_k \leq n }
\left[
\prod\limits_{j \in \{i_1, \cdots, i_k\}}
\left( \frac{\sigma_{j} y_{j}}{2} \right)
\prod\limits_{j \notin \{i_1, \cdots, i_k\}} \left(
1-\frac{\mu_j}{2}
\right)
\right]
\end{equation}
Let us assume that $(y_{i_1}, y_{i_2}, \cdots, y_{i_k})$, 
a $k$-dimensional zero-mean random variable, 
can be characterized using a zero-mean multivariate normal.  Examining the odd terms (ie. the terms $\chi_k$ where $k = 2m+1$), we see that since $\mathds{E}(y_{i_1}y_{i_2} \cdots y_{i_k}) = \mathds{E}\left((-y_{i_1}) (-y_{i_2}) \cdots  (-y_{i_k})\right)=-\mathds{E}(y_{i_1}y_{i_2} \cdots y_{i_k})$, it implies that $\mathds{E}(y_{i_1}y_{i_2} \cdots y_{i_k}) = 0$. Thus, all the odd terms in the expansion of $\braket{\mathcal{O}}_\xrm(t) = \sum\limits_{k=1}^{n+1} \mathds{E}(\chi_k)$ are zero. We will use Isserlis' theorem to simplify the even terms. 
%
Isserlis' theorem states that: 
$
\mathds{E}\left(\prod\limits_{i=1}^k y_i \right) = \sum\prod \mathds{E}(y_i y_j)
$ 
where the product is over the partitions and the sum is over all possible pairwise partitions. When the inter-qubits correlations do not vary across the chip, we get:
$$
\mathds{E}\left(\prod\limits_{i=1}^k y_i \right) = \frac{k!}{2^{k/2} (k/2)!} \rho^k
$$
since the number of ways $(y_1, \cdots, y_k)$ (where $k = 2m$) can be partitioned into $m$ pairs is $(2m)!/ (2^m m!)$. This is because, the first pair can be selected in $2m (2m-1)/2$ ways, the second pair can be selected in $(2m-2)(2m-3)/2$ ways and so on. But the product needs to be divided by $m!$ because the ordering of the $m$ pairs do not matter.

Since the odd-terms are all zero, the mean of the observable can be written as:
\begin{equation}
\begin{split}
\braket{\mathcal{O}}_\xrm(t) =& \mathds{E}\left[ \prod\limits_{i=1}^n \left( 1 - \frac{\xrm_i}{2}\right)\right]= \sum\limits_{k=1}^{n+1} \mathds{E}(\chi_k)\\
=& \sum\limits_{k=1}^{n+1} \left[
(-1)^{k-1} \left( \frac{\sigma_t}{2} \right)^k 
\left( 1 - \frac{\mu_0}{2} \right)^{n-k} \sum\limits_{1 \leq i_1 < i_2 \cdots < i_k \leq n} \mathds{E} ( y_{i_1}y_{i_2} \cdots y_{i_k} )
\right]\\
=& \sum\limits_{m=0}^M 
\left(1-\frac{\mu_0}{2}\right)^{n-2m}
\left( \frac{\sigma_t}{2} \right)^{2m}
\left[
{2M \choose 2m}
\frac{(2m)!}{2^m m!}
\rho^{2m}
\right]\\
=& n! \left[1-\frac{\mu_t}{2}\right]^n\sum\limits_{m=0}^M 
 \frac{ \theta_t^m}{m! (n-2m)!}\\
\sim&  \sqrt{2\pi n}(e/n)^n \left[1-\frac{\mu_t}{2}\right]^n\sum\limits_{m=0}^M 
 \frac{ \theta_t^m}{m! (n-2m)!} \text{ (using Stirling's approximation) }\\
\end{split}
\end{equation}
where
$
\theta_t = \left[
\sigma_t \rho_t/
(1-\mu_t/2)
\right]^2/8
$
and $\mu_t, \sigma_t$ and $\rho_t$ are the time-varying mean, standard-deviation and inter-qubit correlation respectively for the depolarizing noise parameter	 $\xrm$.

\section{Dimensionality reduction and Monte-Carlo sampling }\label{sec:mcmc_appendix}
Sampling from a correlated 16-dimensional time-varying joint density has a very high Monte Carlo sampling overhead (beyond the capacity of our local machines) to achieve convergence in Hellinger distance using Eqn.~\ref{eq:mcmc_hellinger}. However, in absence of correlations, the number of dimensions can be high (at least as high as 16) as sampling from 16 univariate distributions is straightforward and the Hellinger distance converges with low sampling overhead. 

To deal with this problem, as a workaround, we reduce the effective dimensionality using clustering. This is achieved by introducing a minimum threshold for the correlation parameter (any value below that threshold is set to zero).

The threshold is determined by taking into account the computational resources available. A slow machine requires a high threshold while a machine with infinite computing power has threshold as zero (i.e. no dimensionality reduction required). 

Setting a runtime budget of six hours, we deduced that our machine's configuration permitted an effective dimensionality not exceeding $7$ for Monte Carlo convergence. 
To determine the desired threshold for our specific experimental dataset, we run a loop for the correlation threshold from 1.0 to 0.0 in step sizes of -0.01, computing the effective dimensionality (i.e. number of independent clusters) for each month. 
Once we hit the maximum permissible dimensionality, we break out of the loop and choose that as our threshold (0.78 in our study). 

After this, the Hellinger distance computation proceeds as follows. 
Let the noise parameters $(\xrm_0, \xrm_1 \cdots \xrm_d)$ (d=16 in our study), be distributed amongst $K(t)$ independent clusters at time $t$. The variables that belong to a specific cluster are correlated with a correlation coefficient that is above the threshold. 
Let the $i$-th cluster be denoted by $\mathcal{B}_i(t)$. 
Also, let the cardinality of $\mathcal{B}_i(t)$ be denoted by $m_i(t)$. 
It should be clear that: 
$$\sum_i m_i(t) = d, \;\;\;\; \forall t.$$
Let the noise parameters belonging to $\mathcal{B}_i(t)$ be given by $\{ \xrm_{(1,i)}, \cdots, \xrm_{(m_i(t), i)} \}$. 
Finally, let $\Theta_i(t)$ from Eqn.~\ref{eq:copulas} denote the copula function for cluster $\mathcal{B}_i(t)$, i.e., 
\begin{equation}
\Theta_i(t) = \Theta\left[
F_{X_{(1     ,i)}} \left( \xrm_{(1     ,i)};t \right), 
\cdots, 
F_{X_{(m_i(t),i)}} \left( \xrm_{(m_i(t),i)};t \right) 
\right].
\end{equation}
Then, it follows from Eqn.~\ref{eq:mcmc_hellinger} that:
\begin{equation}
\begin{aligned}
1-H^2_{X} =&
\mathds{E}\left( \sqrt{
\frac{ \underset{i \in t_1\text{ clusters}}{\prod} \Theta_i(t_2)}
{\underset{j \in t_2\text{ clusters}}{\prod} \Theta_j(t_1)}
\prod\limits_{k=1}^{d}
\frac{f_{X_k}(\xrm_k;t_2)}
{f_{X_k}(\xrm_k;t_1)}
}\right),
\end{aligned}
\label{eq:time_varying_copula}
\end{equation}

In our experimental dataset, the correlation structure between the noise parameters changes every month. 
Consequently, the number of clusters and their composition changes very month. 
For example, {in May 2022, our methods identified 13 clusters with the biggest cluster comprising 3 noise parameters, while in April 2023, we found 16 independent clusters.} 

\begin{table*}[htbp]
\begin{center}
\scalebox{0.8}{
\small
\begin{tabular}{|l|c|c|c|c|c|c|c|c|c|c|c|c|c|c|c|c|c|c|c|}
\hline
Month
& $H_{X_0}$
& $H_{X_1}$
& $H_{X_2}$
& $H_{X_3}$
& $H_{X_4}$
& $H_{X_5}$
& $H_{X_6}$
& $H_{X_7}$
& $H_{X_8}$
& $H_{X_9}$
& $H_{X_{10}}$
& $H_{X_{11}}$
& $H_{X_{12}}$
& $H_{X_{13}}$
& $H_{X_{14}}$
& $H_{X_{15}}$
& $H_{\text{n}}$
& $H_{\text{a}}$
& $H_{\text{r}}$\\\hline
Jan-22&0.0&0.0&0.0&0.0&0.0&0.0&0.0&0.0&0.0&0.0&0.0&0.0&0.0&0.0&0.0&0.0&0.0&0.0&0.0\\\hline Feb-22&0.82&0.08&0.43&0.38&0.3&0.35&0.28&0.43&0.45&0.39&0.32&0.24&0.62&0.66&0.04&0.26&0.41&0.38&0.971439\\\hline Mar-22&0.97&0.22&0.31&0.3&0.07&0.05&0.31&0.17&0.6&0.22&0.32&0.31&0.11&0.36&0.48&0.17&0.47&0.31&0.99084\\\hline Apr-22&0.64&0.03&0.23&0.53&0.27&0.45&0.21&0.06&0.95&0.11&0.1&0.45&0.33&0.37&0.07&0.61&0.42&0.34&0.978632\\\hline May-22&0.8&0.27&0.61&0.77&0.11&0.4&0.64&0.21&0.38&0.36&0.26&0.65&0.21&0.34&0.34&0.49&0.57&0.43&0.99897\\\hline Jun-22&0.81&0.4&0.43&0.9&0.26&0.34&0.53&0.34&0.28&0.1&0.16&0.36&0.43&0.69&0.16&0.23&0.44&0.4&0.983197\\\hline Jul-22&0.74&0.42&0.96&1.0&0.3&0.44&0.15&0.25&0.48&0.37&0.2&0.16&0.38&0.17&0.14&0.32&0.79&0.41&1.0\\\hline Aug-22&0.89&0.5&0.9&1.0&0.26&0.41&0.53&0.35&0.97&0.31&0.21&0.55&0.22&0.21&0.27&0.3&0.85&0.49&1.0\\\hline Sep-22&0.82&0.48&0.93&1.0&0.45&0.22&0.44&0.34&0.91&0.1&0.08&0.46&0.27&0.31&0.31&0.13&0.79&0.45&1.0\\\hline Oct-22&0.72&0.55&0.9&1.0&0.05&0.42&0.32&0.18&0.95&0.07&0.27&0.43&0.36&0.66&0.74&0.25&0.77&0.49&1.0\\\hline Nov-22&0.36&0.63&0.65&1.0&0.4&0.13&0.55&0.53&0.98&0.22&0.14&0.4&0.29&0.39&0.29&0.3&0.61&0.45&0.999713\\\hline Dec-22&0.42&0.64&0.58&1.0&0.27&0.53&0.45&0.24&0.99&0.17&0.27&0.7&0.65&0.03&0.19&0.37&0.72&0.47&0.999995\\\hline Jan-23&0.46&0.59&0.46&1.0&0.06&0.5&0.31&0.3&0.91&0.53&0.26&0.69&0.55&0.34&0.53&0.46&0.68&0.5&0.999975\\\hline Feb-23&0.45&0.61&0.65&1.0&0.44&0.46&0.44&0.26&1.0&0.12&0.52&0.53&0.4&0.33&0.61&0.34&0.92&0.51&1.0\\\hline Mar-23&0.47&0.5&0.79&1.0&0.22&0.21&0.46&0.33&0.51&0.21&0.4&0.71&0.62&0.09&0.71&0.31&0.64&0.47&0.999876\\\hline Apr-23&0.43&0.55&0.65&1.0&0.16&0.61&0.37&0.25&0.12&0.14&0.15&0.55&0.26&0.42&0.36&0.17&0.61&0.39&0.999721\\\hline
\end{tabular}}
\medskip\medskip\medskip
\caption{Hellinger distance between the distributions of the noise parameters.
}
\Description{This table displays Hellinger distance values for the noise parameters.}
\label{tab:marginal_hellinger19} 
\end{center}
\end{table*}
\vspace{0.5in}
\begin{figure*}[!h]
\centering
\begin{tabular}{ c @{\hspace{40pt}} c }
\includegraphics[width=.4\textwidth]{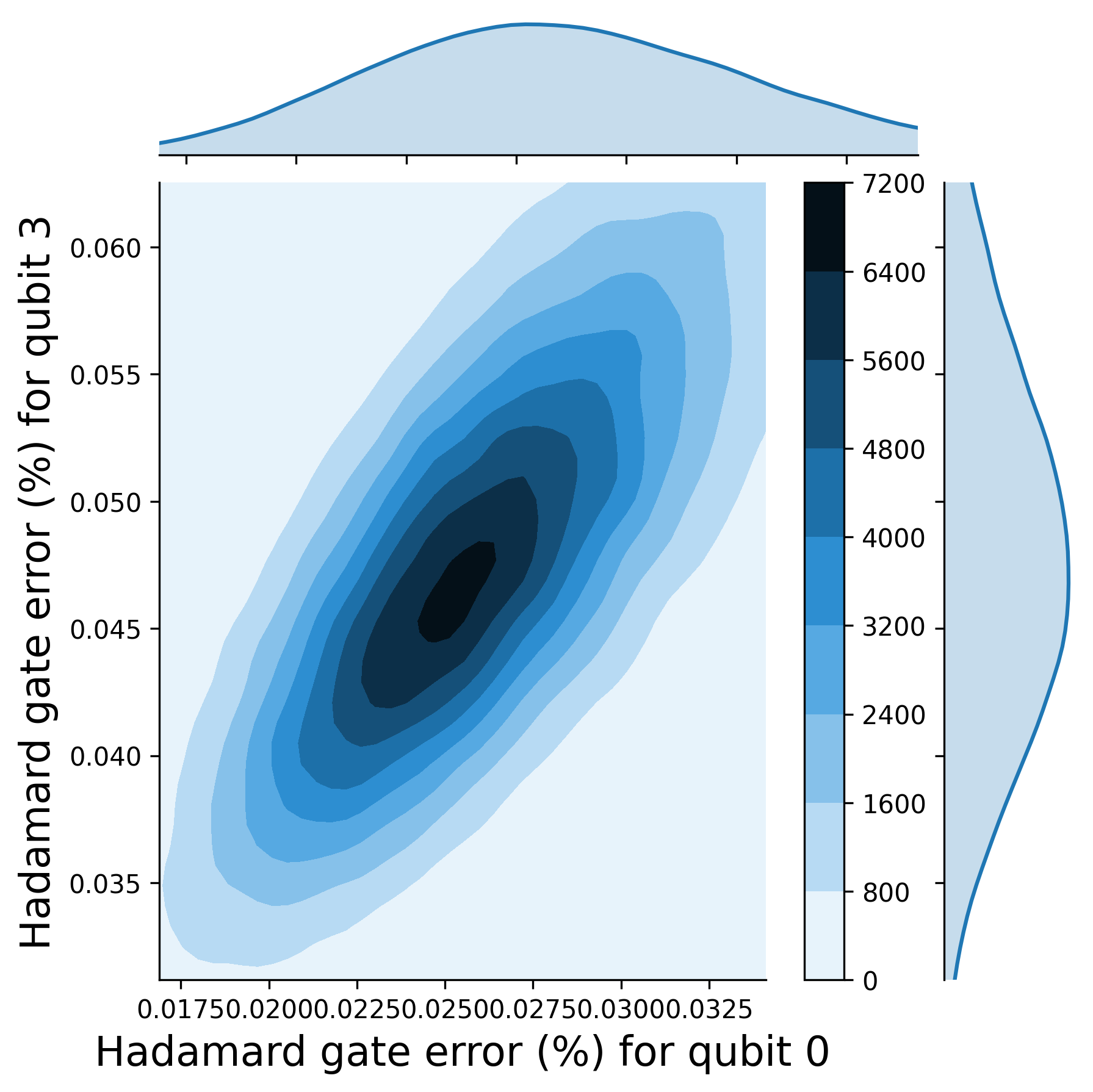}&
\includegraphics[width=.4\textwidth]{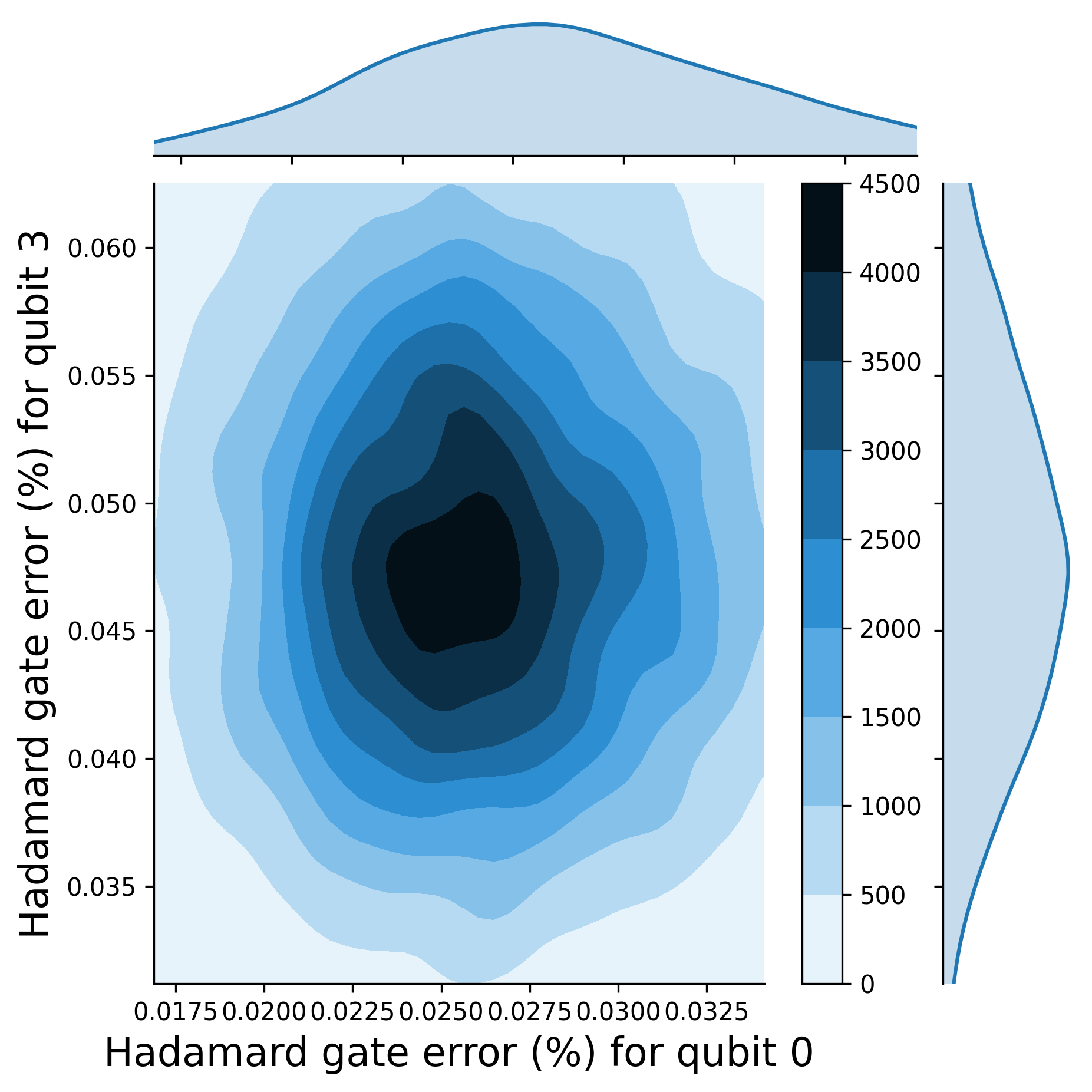}\\
\small (a) & \small (b)\\
\end{tabular}
\medskip
\caption{Visualization of a two-dimensional subset of the Hadamard gate errors for qubit 0 and 3, {which display an 86\% correlation}. We use a contour plot to compare the probability density for Apr-2023 (a) with and (b) without correlation modeling using a copula function. The y-axis represents Hadamard gate error for qubit 3, while the x-axis represents the same for qubit 0. Figure (a) shows an angularly tilted ellipse due to 86\% correlation, while figure (b) shows symmetric concentric circles due to zero correlation. The presence of correlations can significantly change the overlap between the distributions at different times.}
\Description{This graph highlights the importance of correlation modeling for characterizing joint noise distributions for device characterization metrics.}
\label{fig:dist_with_copula}
\end{figure*}
\vspace{0.5in}
\begin{figure*}[!h]
\centering
\includegraphics[width=.8\textwidth]{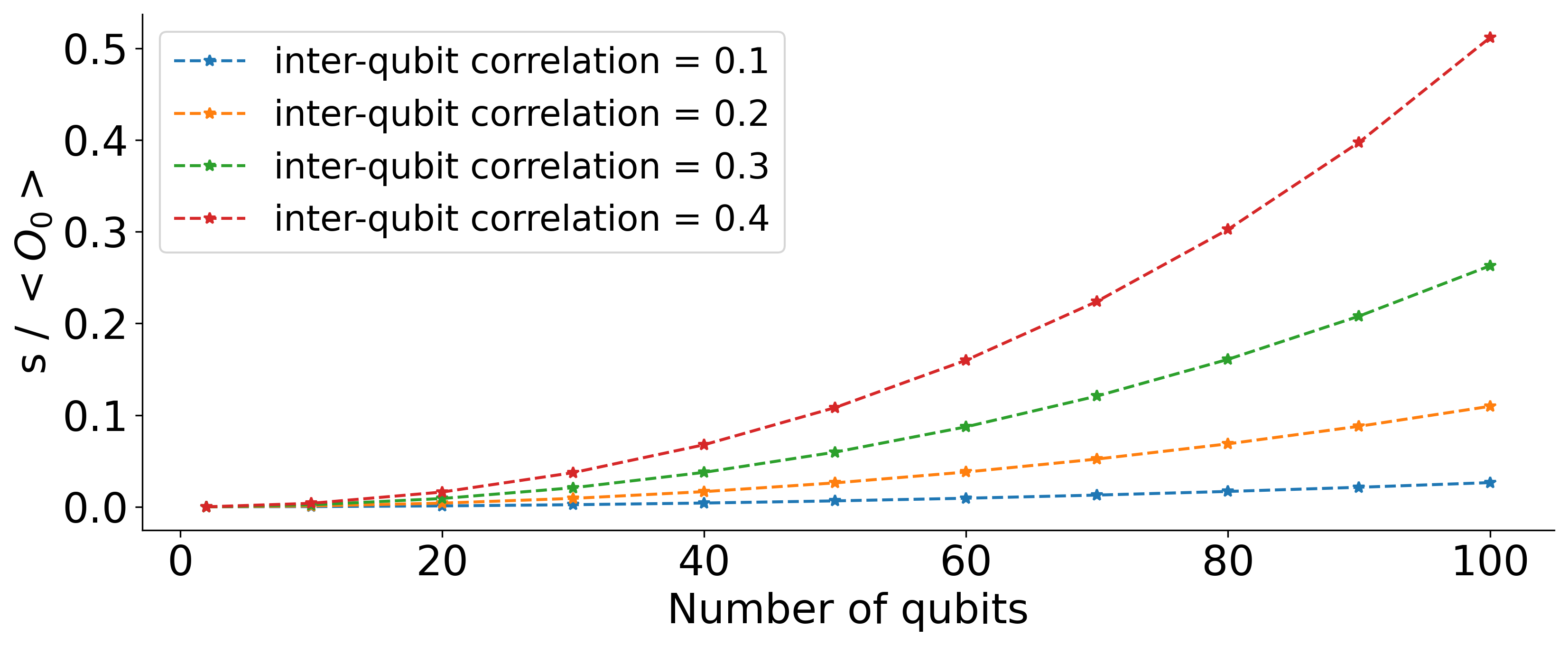}
\medskip
\caption{Outcome stability deteriorates exponentially with number of qubits on an unreliable platform.}
\Description{Outcome stability deteriorates exponentially with number of logical qubits on an unreliable platform.}
\label{fig:depol_sbysmax_large_register}
\end{figure*}
\vspace{0.5in}
\end{document}